\documentclass{emulateapj}
\usepackage{apjfonts}

\shorttitle{Ultradeep $K_S$ Imaging in the GOODS-N}
\shortauthors{Wang et al.}

\begin{document}

\title{
Ultradeep $K_S$ Imaging in the GOODS-N\footnotemark[1]}
\author{Wei-Hao Wang\altaffilmark{2,3}, 
Lennox L. Cowie\altaffilmark{4}, 
Amy J. Barger\altaffilmark{5,6,4}, 
Ryan C.\ Keenan\altaffilmark{5},
and Hsiao-Chiang Ting\altaffilmark{2}}

\footnotetext[1]{Based on observations obtained at the Canada-France-Hawaii 
Telescope (CFHT), which is operated by the National Research Council of 
Canada, the Institut National des Sciences de l'Univers of the Centre 
National de la Recherche Scientifique of France, and the University of Hawaii.}
\altaffiltext{2}{Academia Sinica Institute of Astronomy and Astrophysics, 
P.O. Box 23-141, Taipei 10617, Taiwan}
\altaffiltext{3}{National Radio Astronomy Observatory, 
1003 Lopezville Rd., Socorro, NM, 8701}
\altaffiltext{4}{Institute for Astronomy, University of Hawaii, 
2680 Woodlawn Drive, Honolulu, HI 96822}
\altaffiltext{5}{Department of Astronomy, University of Wisconsin-Madison, 
475 North Charter Street, Madison, WI 53706}
\altaffiltext{6}{Department of Physics and Astronomy, 
University of Hawaii, 2505 Correa Road, Honolulu, H  96822}

\begin{abstract}
We present an ultradeep $K_S$-band image that covers $0.5\times0.5$~deg$^2$ 
centered on the Great Observatories Origins Deep Survey-North (GOODS-N).
The image reaches 
a 5 $\sigma$ depth of $K_{S,\rm AB}=24.45$ in the GOODS-N region, which is 
as deep as the GOODS-N \emph{Spitzer} Infrared Array Camera (IRAC) 3.6~$\mu$m image.  
We present a new method of constructing IRAC catalogs that uses the 
higher spatial resolution $K_S$ image and catalog as priors
and iteratively subtracts fluxes from the IRAC images to 
estimate the IRAC fluxes.  Our iterative method is different from 
the $\chi^2$ approach adopted by other groups.
We verified our results using data taken in two different 
epochs of observations, as well as by comparing our colors with the colors 
of stars and with the colors derived from model spectral energy 
distributions (SEDs) of galaxies at various redshifts.  We make available to 
the community our WIRCam $K_S$-band image and catalog (94951 objects in 
0.25~deg$^2$), the Interactive Data Language (IDL) pipeline used for 
reducing the WIRCam images, and our IRAC 3.6 to 8.0~$\mu$m catalog 
(16950 objects in 0.06~deg$^2$ at 3.6~$\mu$m).  With this improved $K_S$ 
and IRAC catalog and a large spectroscopic sample from our previous work, 
we study the color-magnitude and color-color diagrams of galaxies.  We 
compare the effectiveness of using $K_S$ and IRAC colors to select 
active galactic nuclei (AGNs) and galaxies at various redshifts.  
We also study a color selection of $z=0.65$--1.2 galaxies using the 
$K_S$, 3.6~$\mu$m, and 4.5~$\mu$m bands.
\end{abstract}


\section{Introduction}
Deep near-infrared (NIR) imaging is essential for understanding galaxy 
evolution at high redshifts.  One of its key applications is the 
determination of galactic mass.  NIR luminosity is the best tracer 
of galactic stellar mass because it is less affected by extinction 
and young stars.  
This role is traditionally played by deep ground-based $K$-band imaging 
\citep[e.g.,][]{cowie94,kauffmann98,brinchmann00}.  
An important reason for this is that
the $K$-correction at $K$-band is small across a broad range of 
redshifts and is relatively invariant against galaxy types compared
to the other two NIR bands ($J$ and $H$,
see, e.g., \citealp{keenan09}).  
Recently, the high sensitivity of the
\emph{Spitzer} Infrared Array Camera (IRAC, \citealp{fazio04}) 
adds additional powerful wavebands for this purpose 
\citep[e.g.,][]{fontana06,perez08,elsner08}. However, 
ground-based $K$-band imaging remains important because of the 
ease of observations.

In order to study galaxy formation and evolution, we have been carrying 
out deep NIR imaging  in the Great Observatories Origins Deep Survey-North 
\citep[GOODS-N,][]{giavalisco04}.  In \citet[hereafter B08]{barger08} 
we presented a deep $K_S$-band catalog in the GOODS-N obtained with the 
Wide-field InfraRed Camera (WIRCam) on the 3.6~m Canada-France-Hawaii 
Telescope (CFHT).  The WIRCam image was used by \citet{wang07} to study
an optically faint submillimeter galaxy.  The catalog was used by \citet{cowie08} to 
construct a mass-selected sample of galaxies at $z<1.5$ with an 
AB magnitude limit of 23.4.  We have since added more
data to the WIRCam $K_S$ imaging and slightly improved the reduction.
In this paper we present the new image and a new $K_S$ catalog.  
In order to achieve a more complete 
sampling of galaxy mass at high redshifts, we have also constructed a 
new 3.6 to 8.0~$\mu$m catalog based on \emph{Spitzer} IRAC images in the 
GOODS-N (M.\ Dickinson et al., in preparation), which we also present
here.

Colors in the \emph{Spitzer} IRAC bands are used to study the 
redshifts of galaxies, especially optically faint ones 
\citep[e.g.,][]{pope06,wang07,devlin09,marsden09}.
This is largely based on the rest-frame 1.6~$\mu$m bump in galactic 
spectral energy distributions (SEDs), which is caused by the H$^-$ 
opacity minimum in the stellar photosphere
\citep{simpson99,sawicki02}.  IRAC colors are also used for separating 
active galactic nuclei (AGNs) from galaxies 
(\citealp{lacy04,stern05}, see also \citealp{hatzim05,sajina05}).  
In B08 we studied the effectiveness of using IRAC colors to select 
AGNs and galaxies at various redshifts based on a large 
spectroscopic sample, and we pointed out the limitations in these 
selections.  With our improved color measurements in the IRAC and 
$K_S$ bands, we revisit and discuss these issues.

We describe the observations in \S~\ref{sec_observation}, the 
data reduction in \S~\ref{sec_reduction}, the reduction quality 
in \S~\ref{sec_quality}, and a $K_S$ selected catalog in
\S~\ref{k_catalog}.  We present in \S~\ref{sec_color} a new 
method of constructing IRAC catalogs based on the $K_S$-band 
image and catalog, and we evaluate the performance
of our new method.  In \S~\ref{sec_properties} we describe the 
general properties of the $K_S$ and IRAC catalog, including its 
overlap with other multiwavelength catalogs, the $K_S$ and IRAC 
color-magnitude diagrams and color-color diagrams.  We summarize 
our results in \S~\ref{sec_summary}.  The images and catalog data 
from this work will be available online. 
All fluxes in this paper are $f_\nu$.  All magnitudes are in AB 
system, unless otherwise stated, where an AB magnitude is defined as 
$\rm AB=23.9-2.5\log(\mathrm{Jy})$.

\section{WIRCam Imaging Observations}
\label{sec_observation}

Ultradeep $K_S$-band imaging observations of the Hubble Deep 
Field-North (HDF-N), including the GOODS-N and flanking fields, 
were carried out with WIRCam on the CFHT.  WIRCam consists of four 
$2048\times2048$ HAWAII2-RG detectors covering a field of view of 
$20\arcmin \times 20\arcmin$ with a $0\farcs3$ pixel scale.  
The observations were made in semesters 2006A, 2007A, and 2008A 
by two groups.  A Canadian group led by Luc Simard obtained 
11.5~hr of integration in 2006A.  Our group based in Hawaii 
obtained 37.9~hr of integration between 2006A and 2008A, 
bringing the total integration to 49.4~hr.  To recover the gap 
between the sensors and bad pixels, all exposures were dithered 
with various offsets ranging from $\pm0\farcm5$ to $\pm1\farcm5$ 
along R.A.\ and Dec.  The dither patterns were changed after each 
semester in order to minimize artifacts caused by flat fielding 
or by sky subtraction.  After each 0.5 to 1~hr of dithered 
observations, the dither centers were offset by up to $\pm6\arcmin$
to further broaden the area coverage and smooth out the distribution 
of integration time.  The observations were carried out by the CFHT 
in queue mode with weather monitoring.  Nearly all of the data 
were taken under photometric conditions.  A small fraction ($<10\%$) 
of the data were taken under thin cirrus, so we have down-weighted 
them (see next section).  Typical seeing for the observations is 
between $0\farcs7$ and $0\farcs8$ (FWHM).  Images taken under 
$>1\arcsec$ seeing were not used in the reduction.

\section{Data Reduction}
\label{sec_reduction}

In order to reduce imaging data taken with WIRCam and similar cameras, 
we have developed an Interactive Data Language (IDL) based reduction
pipeline---SIMPLE Imaging and Mosaicking PipeLinE (SIMPLE)\footnotemark[7]. 
Here we describe the SIMPLE reduction of the WIRCam data. 

\footnotetext[7]{The SIMPLE package is available at \\
http://www.asiaa.sinica.edu.tw/$\sim$whwang/idl/SIMPLE.}

\subsection{Removal of Instrumental Features}

The WIRCam images were reduced chip by chip.  Because of the 
rapidly varying sky color in the NIR, only dithered images 
taken within $\sim0.5$~hr were grouped and reduced together.  
A group of raw images were first corrected for nonlinearity 
in the instrumental response
with correction coefficients provided by the CFHT (C.-H. Yan, 2008, 
priv. comm.).  The images were then normalized, median-combined to 
form a sky flat field, and flattened with that.  In the flattened 
images, objects were detected and masked back in the raw images.  
The masked raw images were normalized again by taking into account 
the locations of the masked pixels and then median-combined 
to form a second sky flat.  This was used as the final flat field 
for flattening the raw images again.  Because there was still some 
variation in sky color within the $\sim0.5$~hr period, the sky flat 
did not always perfectly flatten the images.  To remove the residual 
gradients in the flattened images, objects were detected again 
and masked.  A 5th-degree polynomial surface was fitted to each 
masked image and subtracted from the unmasked image. 

On each WIRCam detector, there is crosstalk between each of the 
32 readout channels ($2048\times64$ pixels each), as well as 
within each of the four video boards (8 channels each).
For every image, SIMPLE median-combined the 32 $2048\times64$ 
stripes and subtracted the combined stripe from each of the 
original ones to remove the 32-channel crosstalk features.  
The 8-channel crosstalk features are relatively weak and only 
appear around bright objects after combining several hours of 
images.  SIMPLE repeated the median combination around only 
bright objects and only within the affected 8 channels, and 
then subtracted the median image to remove the 8-channel features.  
All the median combinations here were performed after detected 
objects were masked.  

The above procedures cleanly removed most of the 32 and 8-channel 
crosstalk.  Very low level 8-channel crosstalk features are still 
visible around the brightest $\sim5$ stars in our field-of-view
after combining the 39~hr of data taken in 2006 and 2007.  This 
is probably caused by the coupling between crosstalk, flat 
field pattern, and sky background structures in the images.  
There is no trivial way to solve this, and we did not attempt to 
further remove this very low level effect.  Thanks to the WIRCam 
instrument team, the crosstalk in 2008 had improved behavior and 
no crosstalk residual effect is observed after the above 
32 and 8-channel removal procedure.

\subsection{Correction for Distortion and Astrometry}

The flattened, sky subtracted, and crosstalk removed images were 
then used for deriving the pointing offsets between the dithered 
images and the distortion function of the optics.  In order to 
obtain accurate offsets, SIMPLE uses the package SExtractor 
\citep{bertin96} to detect objects and to determine their pixel 
coordinates.  Then offsets are derived by averaging the 
displacements of the same objects that appear in different images.  
Because of the optical distortion, objects in different parts of 
the images have slightly different dither displacements.  This 
differential displacement is used by SIMPLE to derive 
the distortion function, following the technique developed by 
\citet{anderson03}.  

SIMPLE uses cubic polynomials to fit the distortion functions, i.e., 
\begin{eqnarray}
x^{\prime} = F(x,y) = \sum f_{ij}x^iy^j, \\   
y^{\prime} = G(x,y) = \sum g_{ij}x^iy^j,
\end{eqnarray}
where $x$ and $y$ are the undistorted coordinates of objects, and 
$x^{\prime}$ and $y^{\prime}$ are the distorted coordinates, $F$ and 
$G$ are the distortion functions, and $i+j\le3$.  To first order, 
the displacements of objects between dithered images can then be 
expressed as the expansions of the distortion functions:
\begin{eqnarray}
\Delta x^{\prime} = \frac{\partial F(x,y)}{\partial x} \Delta x + 
\frac{\partial F(x,y)}{\partial y} \Delta y, \\
\Delta y^{\prime} = \frac{\partial G(x,y)}{\partial x} \Delta x + 
\frac{\partial G(x,y)}{\partial y} \Delta y, 
\end{eqnarray}
where $\Delta x^{\prime}$ and $\Delta y^{\prime}$ are the measured
dither displacements of objects, which are functions of positions, and 
$\Delta x$ and $\Delta y$ are the pointing offsets of the dithering.

Under the above approximation, the measured displacements 
of objects are the first order derivatives of the distortion 
functions.  For $n$ 
objects in $m$ dithered images, there are approximately 
$n\times(m-1)$ sets of linear equations for each of 
$\Delta x^{\prime}$ and $\Delta y^{\prime}$.  The systems of linear
equations were solved for the coefficients of $F$ and $G$ with a
least-squares method.  Initially, the mean values of $\Delta x^{\prime}$ 
and $\Delta y^{\prime}$ were used as rough estimates of $\Delta x$ and 
$\Delta y$.  After $F$ and $G$ were integrated, they were used to
correct for $x^{\prime}$ and $y^{\prime}$ to derive better estimates 
of the dither offsets $\Delta x$ and $\Delta y$.  Then improved coefficients
of $F$ and $G$ were solved with the new $\Delta x$ and $\Delta y$.  
For WIRCam, one such iteration seems to provide sufficiently good 
distortion functions. Since the optical distortion was derived 
within a $\sim0.5$ hr sequence of dithered images, the above approach 
has the advantage of accounting for the long-term time-dependent 
flexure of the telescope.  It also does not require any
external information and thus can easily be applied to any target field.

To determine the absolute astrometry, we used several different 
catalogs in the HDF-N region.  We first combined the catalogs of 
the Sloan Digital Sky Survey\footnotemark[8] (SDSS) Data Release~6
around the HDF-N region
and the v1.0 GOODS-N Advanced Camera for Surveys (ACS) catalog 
\citep{giavalisco04}, after correcting for the well-known $0\farcs38$ 
offset between the GOODS-N ACS frame and the radio frame.  The combined 
catalog was cross-checked with the source positions in the Very Large 
Array (VLA) 1.4~GHz catalog of \citet{biggs06}
to ensure consistency with the 
radio frame.  The comparison between the ACS and the VLA astrometry 
is only good to several tens of milli-arcseconds, because of the 
differences in optical and radio morphologies of distant galaxies.
This sets the limit for our absolute astrometry.

We then reduced several hours of the best quality 
WIRCam $K_S$-band data and matched the WIRCam object positions
to the above SDSS/ACS astrometric catalog.  We added bright and 
compact WIRCam $K_S$ objects to the SDSS/ACS catalog to form a dense 
astrometric catalog for reducing the full set of WIRCam $K_S$ images.  
The final catalog has $\sim8700$ objects over a diameter of 1.3~degree, 
approximately 5700 of which are concentrated in the central
0.6~deg$^2$ region.  In all the reductions, when individual images were
warped to correct for distortion and projected onto a common sky plane, 
the objects detected in the images were forced to match the positions 
in the astrometric catalog.  The registration of the images onto the 
catalog was performed at the sub-pixel level with a bilinear resampling. 
There was only one resampling for each exposure in 
the entire reduction, which was a fifth degree polynomial including 
both the distortion correction and a tangential sky projection.
This one-step resampling minimized the image smearing.

\footnotetext[8]{Funding for the SDSS and SDSS-II has been provided 
by the Alfred P. Sloan Foundation, the Participating Institutions, 
the National Science Foundation, the U.S. Department of Energy, 
the National Aeronautics and Space Administration, the Japanese 
Monbukagakusho, the Max Planck Society, and the Higher Education 
Funding Council for England. The SDSS Web Site is 
http://www.sdss.org/.\\
The SDSS is managed by the Astrophysical Research Consortium for 
the Participating Institutions. The Participating Institutions are 
the American Museum of Natural History, Astrophysical Institute Potsdam, 
University of Basel, University of Cambridge, Case Western Reserve 
University, University of Chicago, Drexel University, Fermilab, 
the Institute for Advanced Study, the Japan Participation Group, 
Johns Hopkins University, the Joint Institute for Nuclear 
Astrophysics, the Kavli Institute for Particle Astrophysics and 
Cosmology, the Korean Scientist Group, the Chinese Academy of 
Sciences (LAMOST), Los Alamos National Laboratory, the 
Max-Planck-Institute for Astronomy (MPIA), the Max-Planck-Institute 
for Astrophysics (MPA), New Mexico State University, Ohio State 
University, University of Pittsburgh, University of Portsmouth, 
Princeton University, the United States Naval Observatory, and 
the University of Washington.}

\subsection{Mosaicking and Photometric Calibration}

Before the distortion-corrected images were combined to form a 
deep mosaic, the images were calibrated for extinction variation, 
cosmic rays were removed, and the images were weighted.  
The relative extinction 
variation was corrected using fluxes measured by SExtractor on 
high S/N objects in the images.  Then pixels in different images 
that have the same sky coordinates were compared for rejecting 
cosmic rays with sigma clipping.
The clipping rate was kept to be approximately 1\% so only an 
insignificant amount of useful data were discarded.  The sigma 
clipping allows weighted mean combinations 
of the images, which produces 25\% higher S/N compared to 
median combinations when all images have equal weights.
The images were optimally weighted according to the atmospheric 
transparency ($\eta_{\rm atm}$, derived from the flux calibration, see below), 
integration time ($T_{\rm exp}$), and background brightness ($BG$).  
Furthermore, each pixel was weighted by its relative efficiency 
($Q_{\rm E}$, proportional to flat field).   Thus the
final weight applied to a pixel is:
\begin{eqnarray}
w(x,y) \propto \eta_{\rm atm}^2 T_{\rm exp} BG^{-1} Q_{\rm E}(x,y).
\end{eqnarray}\label{weight_eq}
Since objects in the images have a wide range of sizes, we did 
not weight the images with seeing.
The relative extinction corrected, cosmic ray removed, and
weighted images were then mean-combined to form a deep mosaic.

\begin{figure}
\epsscale{0.95}
\plotone{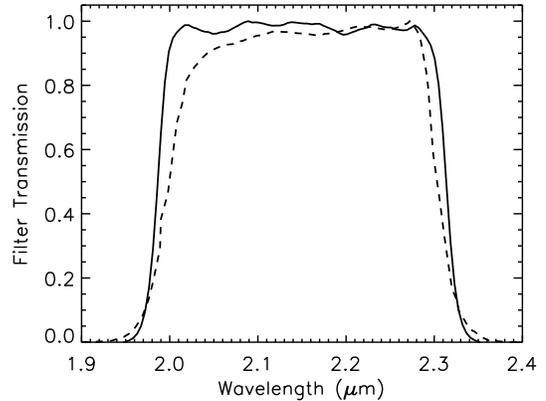}
\caption{$K_S$ filter transmission curves.
The \emph{solid} curve is the WIRCam $K_S$ filter and the
\emph{dashed} curve is the 2MASS $K_S$ filter.
\label{fig_filter}}
\end{figure}

On the mosaicked images, object fluxes were measured again with 
SExtractor using $5\arcsec$ diameter apertures, and the results were 
cross-correlated with the Two Micron All Sky Survey\footnotemark[9]  
(2MASS, \citealp{2mass}) point-source catalog.  
We used 2MASS point sources with $K_S=14.44$--16.34 to calibrate the 
absolute flux of the WIRCam image.  Objects brighter than this 
range are affected by WIRCam nonlinearity (despite the correction 
for linearity) and objects fainter than this range
show a significant selection effect in the 2MASS fluxes.   
We adopted the 2MASS ``default magnitudes,'' which attempt to account for
the total fluxes of the point sources \citep{cutri03}.
Because the 2MASS $K_S$ and WIRCam $K_S$ filter transmission 
curves are very similar (Figure~\ref{fig_filter}), we directly adopted the 2MASS 
$K_S$ magnitudes and did not attempt to account for the slight difference 
between the two systems.  We adopted the 666.7~Jy 
$K_S$ Vega zero-magnitude provided by 2MASS \citep{cohen03}.  
Calibrated mosaic images from all the dither sets and from the 
four chips were combined into a final wide-field, ultradeep image.  
All the weights described above were carried into the final image 
in the form of effective integration time.  We found that
despite the large variation in sky background brightness, noise 
in the final image scales nicely with the inverse square-root 
of the effective integration time in intermediate and shallow 
regions where confusion noise is less important.
This means that our weighting scheme provides excellent S/N 
optimization.

\footnotetext[9]{2MASS is a joint project of the University of 
Massachusetts and the Infrared Processing and Analysis 
Center/California Institute of Technology, funded by 
the National Aeronautics and Space Administration and the 
National Science Foundation.}

\begin{figure}
\epsscale{1.0}
\plotone{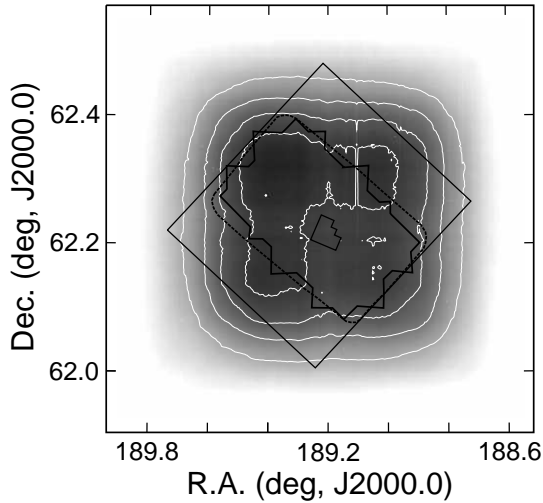}
\caption{Survey field layout.  Grayscale shows the effective 
integration time of our WIRCam $K_S$ imaging. White contours
show 80\%, 60\%, 40\% and 20\% of the maximum effective integration time.
Black solid 
polygons show the GOODS-N ACS region and the HDF-N-proper.  
Black dashed contour shows the GOODS-N IRAC region.  Black 
rectangle shows the \emph{Chandra} Deep Field-North region.
\label{fig_layout}}
\end{figure}

\begin{figure}
\epsscale{1.0}
\plotone{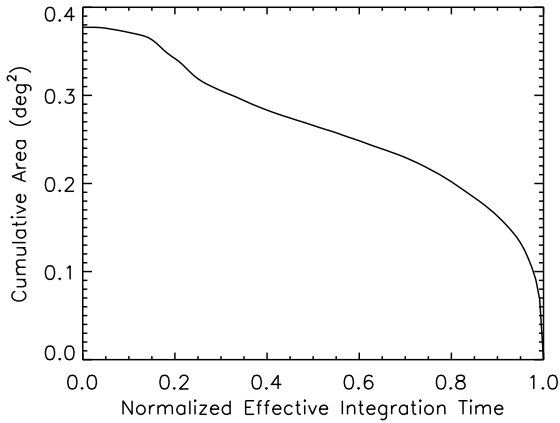}
\caption{Cumulative area in the WIRCam $K_S$ image vs.\ 
effective integration time.  The maximum effective integration time
in the map is normalized to be 1.0.  The effective integration time 
accounts for variation in efficiency of pixels, atmospheric 
extinction, and background brightness (see text).  In shallow 
and intermediate depth regions where confusion is less important, 
noise very closely scales as inverse square-root of the
effective integration time.  
\label{fig_area}}
\end{figure}

Figure~\ref{fig_layout} shows the integration time distribution 
and layout of our survey field. 
Figure~\ref{fig_area} shows the cumulative area in the final 
mosaic as a function of effective integration time. 
The final combined mosaic covers a square area of 
$35\arcmin\times35\arcmin$, centered at
$\alpha=189\arcdeg.227$, $\delta=62\arcdeg.239$ (J2000.0).  
The central $\sim18\arcmin$ region has more than 20~hr 
of integration at each position and fully covers the GOODS-N 
ACS area.  The image has uniform image quality across the 
entire field, with a FWHM of $\sim0\farcs8$.

\section{Reduction Quality}\label{sec_quality}

\subsection{Astrometry}

\begin{figure}[t]
\epsscale{1.0}
\plotone{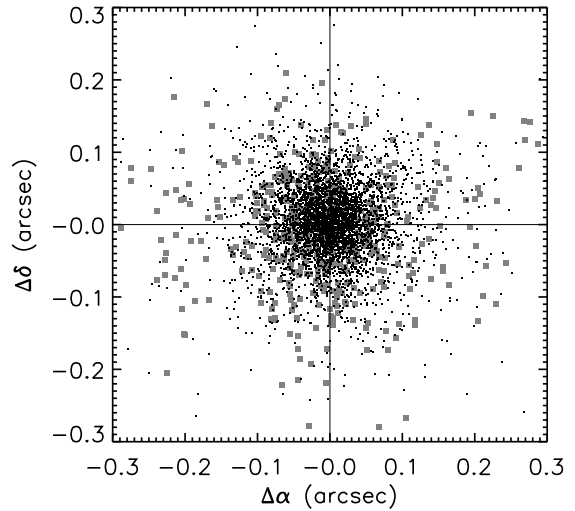}
\caption{Relative astrometric offsets between the WIRCam $K_S$ image
and the SDSS and GOODS-N ACS catalogs.  Approximately 4000
well-detected compact sources (S/N~$>20$, FWHM~$<1\farcs2$) are
plotted.  \emph{Squares} show the offsets of the WIRCam positions
with respect to the SDSS positions, and \emph{dots} show those with 
respect to the ACS positions.
\label{fig_astrometry_1}}
\end{figure}

\begin{figure*}
\epsscale{1.0}
\plotone{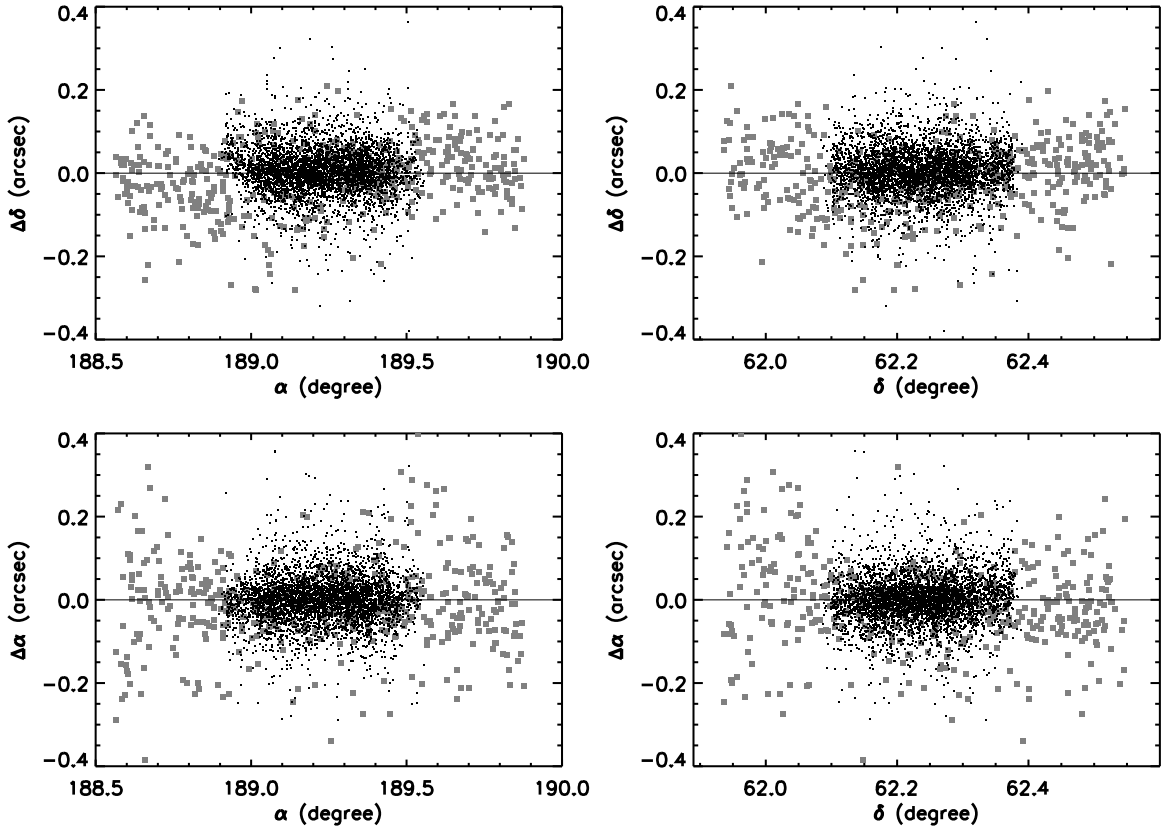}
\caption{Same as Figure~\ref{fig_astrometry_1}, but as functions 
of RA and Dec. 
\label{fig_astrometry_2}}
\end{figure*}

The SIMPLE reduction of the WIRCam $K_S$ image forced the 
positions of detected objects to match those in the astrometric 
catalog formed with the SDSS and the GOODS-N ACS catalogs.  
The resultant astrometry is excellently constrained by the ACS 
catalog in the central region and reasonably constrained by the 
SDSS catalog in the outer region.  To quantify the 
(relative) astrometric accuracy, we selected well-detected 
compact sources (S/N~$>20$ and FWHM~$<1\farcs2$) from the 
SExtractor catalog (Section~\ref{k_catalog}).  Nearly 4000 such 
sources have counterparts in the SDSS or GOODS-N ACS catalog.  
Figure~\ref{fig_astrometry_1} and Figure~\ref{fig_astrometry_2} 
show the offsets between the WIRCam and the SDSS/ACS positions.
There are no observable systematic offsets between the WIRCam positions
and both the SDSS and ACS catalogs.  However, the astrometry is clearly 
better in the central part of the image where ACS sources were used 
to constrain the astrometric solution.  The rms dispersions of the offsets 
are $0\farcs05$ between WIRCam and ACS along both the R.A.\ and Dec.\ axes on
nearly 4000 sources.  The rms offsets between WIRCam and SDSS are larger,
$0\farcs08$ for RA and $0\farcs10$ for Dec, on nearly 460 sources.
We found that the SDSS and ACS catalogs have similar $\sim0\farcs1$ rms 
offsets between each other.  This suggests the larger observed offsets 
between the WIRCam and SDSS positions may primarily be a consequence
of an uncertainty internal to the SDSS catalog.

The above measured rms offsets include both systematic and random
errors.  For example, the measured $0\farcs05$~rms offset between
WIRCam and ACS is slightly larger than the nominal positional uncertainty
for FWHM~$\sim0\farcs8$ and S/N~$>20$, which should be $<0\farcs04$.
To estimate the systematic error in astrometry, we repeated the same
measurements with higher S/N cuts and lower FWHM cuts.
The rms offset decreases initially but stops decreasing when 
it approaches $0\farcs03$, even when the selected sources have 
extremely high S/N (e.g., $>200$) and have FWHM comparable to the seeing.
We conclude that there is an $\sim0\farcs03$ systematic
uncertainty in our WIRCam astrometry in the GOODS-N ACS region.  
Such an error includes both the internal error in the ACS catalog 
and the error introduced by the image registration of SIMPLE.  On 
the other hand, the rms offset between WIRCam and SDSS is $\sim0\farcs1$.  
Since all SDSS sources are detected in the WIRCam image with S/N~$>100$
(median S/N is 2300), random noise is negligible here.  We conclude
that the systematic error is $\sim0\farcs1$ in outer regions where 
ACS positions are unavailable.

\subsection{Photometry}

\begin{figure}
\epsscale{1.0}
\plotone{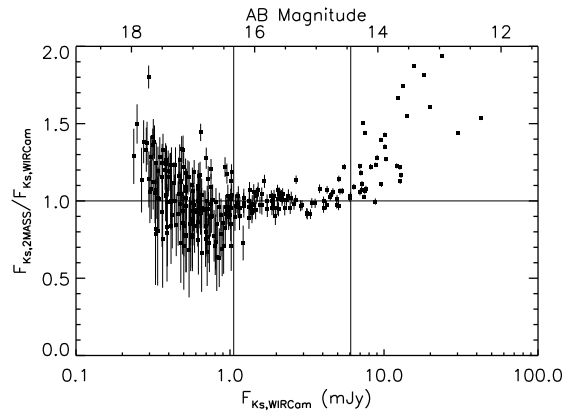}
\caption{Comparison of the 2MASS $K_S$ fluxes with the 
calibrated WIRCam $K_S$ fluxes.  The noise in the WIRCam fluxes is 
plotted as horizontal error bars but is too small to be visible for the entire 
flux range and therefore makes a negligible contribution to the vertical error bars.
This implies a calibration limited by the photometric quality of 2MASS.
Vertical lines show the flux range used for the calibration.  Sources outside
the two lines are not used because of nonlinearity in WIRCam fluxes and 
selection effects in 2MASS.  Within the two vertical lines, data that are 
$>3\sigma$ away from the mean are also excluded in the calibration. 
\label{fig_photometry_1}}
\end{figure}

The absolute flux scale of our WIRCam $K_S$ image was calibrated with 
2MASS point sources in the image.  Figure~\ref{fig_photometry_1} shows 
the comparison between the 2MASS fluxes and the calibrated WIRCam fluxes
derived with $5\arcsec$ diameter apertures.  The ratio between the two
is only flat in a narrow flux range.  We chose $K_{s,AB}=14.44$--16.34
(the two vertical lines in Figure~\ref{fig_photometry_1}) for the 
calibration of the WIRCam image. Outside this range, there are 
nonlinearity problems with the WIRCam
sensors (bright end) and selection effects in 2MASS (faint end).  
In the chosen range for calibration, the error-weighted mean of the 
flux ratios ($F_{\rm 2MASS}/F_{\rm WIRCam}$) for 70 sources is 
$1.0005 \pm 0.5\%$, indicating that 
the overall absolute calibration of the WIRCam $K_S$ image is good 
to $\sim0.5\%$.  We note that the error bars in 
Figure~\ref{fig_photometry_1} are dominated by 2MASS flux errors.
Thus, the calibration of our WIRCam image is limited by the photometric
quality of 2MASS.  It is not possible to achieve better results unless 
we make frequent monitoring of standard stars during our observations, 
which is quite expensive in terms of observing time.

We also investigated the photometric uniformity in our WIRCam $K_S$ 
image.  The above value of 0.5\% is the mean calibration 
error over the entire image. It is possible that the calibration of 
certain smaller regions is worse than this value.  We subdivided the 
WIRCam image into four quadrants, roughly corresponding to data obtained 
from the four detector chips, and then compared the WIRCam fluxes 
with the 2MASS fluxes as we did for the entire image.
For quadrants 1 through 4, the error-weighted flux ratios are
$1.010\pm0.9\%$ (16 sources), $-0.985\pm0.9\%$ (19 sources), 
$1.009\pm1.3\%$ (16 sources), and $1.011\pm0.9\%$ (22 sources), respectively.
In all of these cases, based on the numbers of available 2MASS sources and 
their 2MASS flux errors, the expected calibration errors are roughly 1\%.
The standard deviation of the four measured offsets is 1.3\%, which 
compares well
with the expected calibration error from 2MASS.  This suggests that 
there is not an observable photometric gradient over size scales of 
$\sim15\arcmin$.  The absolute fluxes of objects are still limited by the
photometric quality of 2MASS at this scale.  The measured 1.3\%
variation is indeed excellent when compared to other $K$-band 
extragalactic deep surveys over similar size scales 
\citep[e.g., 2\% in the COSMOS survey,][]{capak07}.

\section{$K_S$ Source Catalog}
\label{k_catalog}

We used SExtractor (v2.5.0) to produce a $K_S$ source catalog.  
The most important source extraction parameters in SExtractor are 
summarized in Table~\ref{tab_sex}.  We provide our raw SExtractor 
$K_S$ catalog in this data release.  We caution that the detection parameters 
(the first three in the table) that we adopted are somewhat aggressive in
detecting faint sources, which may lead to more spurious sources 
(see \S~\ref{sec_spurious}).
The users of our data product may want to regenerate catalogs with
more conservative values, based on the nature of their work.

Based on the raw catalog, we compiled a $K_S$ selected catalog, 
which is the primary product of this paper.  We first estimated 
the true noise level in the WIRCam $K_S$ image.  We masked the detected 
objects and convolved the image with circular kernels of amplitude 1.0 
(i.e., flux apertures).  The local (within $\sim14\arcsec$) rms 
fluctuations in the convolved image were considered as errors in 
aperture photometry that include all possible sources of noise: 
photon noise, 
correlated noise between pixels, confusion noise from faint, undetected 
objects, and uncertainty in the estimate of the small-scale background.  
We compared the noise measured in this way with the photometric errors 
derived by SExtractor.  We found that for all aperture sizes between
$1\farcs5$ and $5\arcsec$ in diameter, the ratio between the two 
errors is fairly insensitive to the aperture size and has a value of 
1.28.  Thus, we multiplied all of the flux errors in the
SExtractor catalog by this factor to approximate the total noise.  
This does not affect the flux measurements.

\begin{deluxetable}{lr}
\tablewidth{0pt}
\tablecaption{SExtractor Parameters \label{tab_sex}}
\tablehead{\colhead{Parameter} & \colhead{Value}}
\startdata
DETECT\_MINAREA 	& 2 \\
DETECT\_THRESH    	& 1.25 \\
ANALYSIS\_THRESH  	& 1.25 \\
FILTER           			& Y \\
FILTER\_NAME		& gauss\_1.5\_3x3.conv \\
DEBLEND\_NTHRESH  	& 64 \\
DEBLEND\_MINCONT  	& 0.00001 \\
CLEAN            			& Y \\
CLEAN\_PARAM      	& 0.1 \\
SEEING\_FWHM      	& 0.8 \\
BACK\_SIZE        		& 24 \\
BACK\_FILTERSIZE  	& 6 \\
BACK\_TYPE        		& AUTO \\
BACKPHOTO\_TYPE   	& LOCAL \\
BACKPHOTO\_THICK  	& 24 \\
WEIGHT\_TYPE      	& MAP\_WEIGHT \\
WEIGHT\_THRESH    	& 10 
\enddata
\end{deluxetable}

The next step is to estimate aperture corrections.  In our data 
reduction we calibrated the flux on 2MASS point sources using fixed 
$5\arcsec$ diameter apertures.  
Thus, SExtractor fluxes measured with $5\arcsec$ apertures do not
need any aperture corrections and 
we consider $5\arcsec$ aperture fluxes to be the best 
estimate of the total fluxes of the sources.  
This clearly does not apply to the several most extended objects 
in the image, since their aperture corrections should be much larger 
than that for smaller objects (i.e., 2MASS point sources used in the
calibration). Since they are not the major targets of our 
work, we did not attempt to obtain highly accurate fluxes for these very few sources.  

Despite that $5\arcsec$ aperture fluxes are the best estimate of
total fluxes, their S/N is relatively low.  We thus looked for suitable fluxes measured
with smaller apertures, and corrected the small aperture fluxes to $5\arcsec$ when it is necessary.
We compared all other SExtractor flux options with the fixed
$5\arcsec$ diameter aperture fluxes and found that the SExtractor 
``auto-aperture'' fluxes agree excellently with them at all flux levels.  
In addition, the auto-aperture fluxes have the best optimized 
S/N for sources with various morphologies and fluxes.  We thus 
decided to adopt directly the SExtractor auto-aperture fluxes and 
errors in our final catalog.  We did not apply any  additional corrections
except for the aforementioned factor of 1.28 to the errors.

The adoption of SExtractor auto-aperture fluxes works excellently for nearly
all sources.  However, a very small number of sources (0.7\% of total)
are undetected in auto-apertures but detected in fixed apertures.
For such sources, we adopted their $3\arcsec$ fluxes and corrected the fluxes to 
$5\arcsec$ with a median correction factor from all other detected sources.
In addition, after all the above corrections to the fluxes and flux errors, 
some sources in the SExtractor raw catalog are still undetected.
Given that our error estimate is realistic and the 
flux is well optimized, we considered sources with S/N~$<1$ as 
undetected (1.9\% of total) and only included their flux errors in the final catalog 
for upper limits.  To keep the source number and order in our final $K_S$ 
catalog identical to those in the SExtractor raw catalog, 
we did not remove such undetected sources.
This is for easy comparisons with the SExtractor catalog, 
in case the users of our data need to carry out more consistency or quality check.

We next evaluated the depth and completeness of our $K_S$ WIRCam catalog.
In the central $30\arcmin\times30\arcmin$ of the image where the 
sensitivity distribution is more uniform, the median 1$\sigma$ error 
among 3$\sigma$ sources is 0.15~$\mu$Jy, corresponding to a 
3$\sigma$ limiting magnitude of 24.77.
In the GOODS-N region, where the image is the deepest,
the median 1$\sigma$ error is 0.12~$\mu$Jy, corresponding to a 3$\sigma$ 
limiting magnitude of 25.10.  This is as deep as the IRAC 3.6~$\mu$m image 
in the same region.

\subsection{Depth and Completeness}

We estimated the completeness of our $K_S$ catalog with Monte Carlo 
simulations.  
We note that completeness is a complex function of flux, 
morphology, crowdedness, and source extraction.  The results here are 
meant only to provide a guide.  Exactly how one should carry out the 
completeness computation depends on the purpose of the study, and 
we shall leave this to the users of our data.

\begin{figure}
\epsscale{1.0}
\plotone{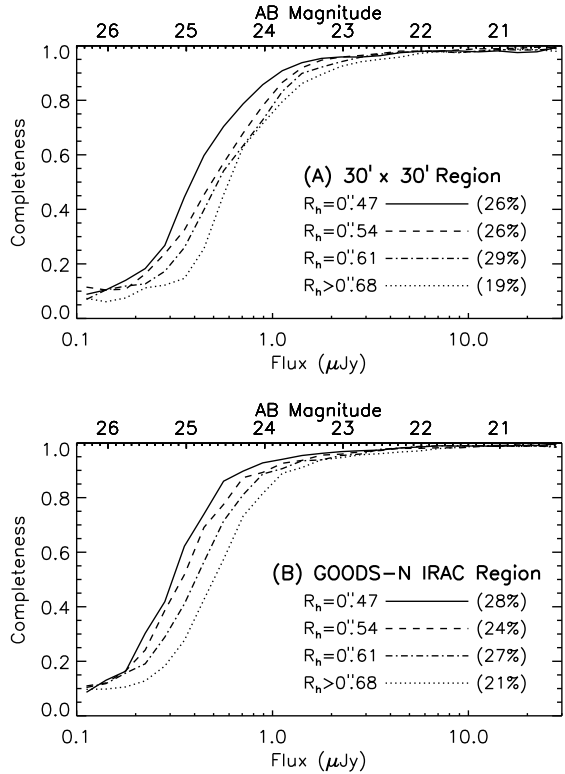}
\caption{Completeness of the $K_S$ WIRCam catalog vs. flux and 
half-light radius in the central $30\arcmin \times 30\arcmin$ region (A)
and in the GOODS-N IRAC region (B).  Percentages given in the parentheses are 
fractions of objects in the four half-light radius groups in the $K_S$ catalog.
\label{fig_completeness}}
\end{figure}

Here, we first randomly selected well-detected (S/N~$>20$) sources from the 
SExtractor catalog. The selected sources span a wide range of flux, 
size, and shape. We extracted these sources from the WIRCam image and 
artificially dimmed or brightened them to the flux range of interest. 
Then we randomly placed them in the WIRCam image and ran 
SExtractor to detect them.  Each time we only placed a small number 
of sources in the image so the overall source density did not change 
by $\gtrsim1\%$. A total of $\sim50,000$ sources were randomly placed 
in the image. The SExtractor catalogs for simulated images 
were processed identically to the real catalog.  We considered a 
source recovered if it is detected at $>2\sigma$.  We divided sources 
into four groups according to their half-light radii
($R_h$) and calculated the recovery rate in each 0.1 dex flux bin 
and $R_h$ group.  The results in the central $30\arcmin\times30\arcmin$ region
are shown in Figure~\ref{fig_completeness}a. 
The 90\% completeness limits for the two compact groups 
($R_h \sim 0\farcs47$ and $0\farcs54$, comprising $>50\%$ of 
the total sample) are 1.10~$\mu$Jy ($7.3\sigma$) and 1.32~$\mu$Jy ($8.8\sigma$), 
respectively, equivalent to 23.80 and 23.60 magnitude.  
For the entire sample the mean 90\% 
completeness weighted by the size distribution is 1.38~$\mu$Jy ($9.2\sigma$, 
which is equivalent to 23.55 magnitude.  If we restrict this to 
the deeper 0.06~deg$^2$ GOODS-N IRAC region 
(Figure~\ref{fig_completeness}b), then the mean 90\% completeness 
limit becomes 0.94~$\mu$Jy ($7.8\sigma$), or 23.96 magnitude.

The above results show that the present WIRCam imaging is the deepest 
in the $K$-band over a substantial area.  For example, the 
Very Large Telescope (VLT) NIR imaging in the GOODS-S is 90\% complete 
at $K_S=23.8$ in a small area of $<10$~arcmin$^2$ \citep{grazian06}.  
On the other hand, our team \citep{wang09} and Japanese groups 
\citep{kajisawa06} have carried out extremely deep $K_S$ band imaging 
in the GOODS-N at the 8.2~m Subaru Telescope with a smaller 
$2048 \times 2048 \times 2$ imager.  \citet{kajisawa09} combined 
all archived data to make a deep $K_S$ image of the GOODS-N.  They 
achieved 3$\sigma$ limiting magnitudes of 25.5 and 26.5 in 
103~arcmin$^2$ and 28~arcmin$^2$, respectively.  This Subaru image 
is deeper than our WIRCam image (3$\sigma$ of 25.1
magnitude in the $\sim200$ arcmin$^2$ GOODS-N region), but the area 
coverage is much smaller.

\subsection{Spurious Sources}\label{sec_spurious}

We took two approaches to estimate the spurious fraction, one based on
an inverted image and the other based on multi-wavelength counterpart identification.
First, we constructed an image with inverted background, and then ran through 
the same source detection and catalog compilation described above.  
This infers a spurious fraction of $\sim20\%$ on sources with S/N $>3$, which is 
unreasonably high.  After examining the image and the ``sources,'' we believe this is 
because the crosstalk of WIRCam produces negative holes near bright objects, 
making the noise non-Gaussian and greatly overestimating the spurious fraction.  
One evidence for this is that, in relatively crosstalk-free regions, the spurious fraction 
inferred from the inverted image dramatically decreases to $5\%$.  This value of 5\% 
is an upper limit since there are still low-level crosstalk effects.  

We then looked for $K_S$ sources that are not detected in other images.  
In the region where the GOODS-N ACS and IRAC images overlap, there
are 15112 $K_S$ sources detected at $3\sigma$.  Among them,
489 (3.2\%) sources do not have counterparts in the ACS catalog and are not detected at 
3.6 and 4.5 $\mu$m at $3\sigma$ in the IRAC catalog described in the
next section.  (See \S~\ref{sec_cat_overlap} for a slightly different comparison.)  
The fraction 3.2\% is also an upper limit since we cannot rule out real $K_S$ sources
being not detected in the ACS and IRAC images, especially given that the IRAC images
are not substantially deeper than the $K_S$ image.  This 3.2\% upper limit
is tighter than the 5\% limit set by the inverted image, consistent with the above
suggested non-Gaussian noise.  Because of this limitation, it is difficult to
further constrain the spurious fraction of our $K_S$ catalog.  We suggest the users
of our data to carry out independent source extraction if more conservative extractions 
 are desired.

In Table~\ref{tab_prop} we summarize the basic properties of our 
WIRCam $K_S$ image described in this and previous sections.

\begin{deluxetable}{lc}
\tablewidth{0pt}
\tablecaption{Properties of the WIRCam Images \label{tab_prop}}
\tablehead{ \colhead{Properties} & \colhead{$K_S$ Image} } 
\startdata
Semesters 				& 2006A, 2007A, 2008A 	\\ 
P.I. 						& L. Cowie, L. Simard 	\\ 
Total Integration 			& 49.4 hr \\
Field Center (J2000.0) 		& \\ 
~~~~R.A.					&$189\arcdeg.227$ \\
~~~~Dec.					& $62\arcdeg.239$\\
Area	 ($T_{exp}>2$ hr)		& 0.25 deg$^2$ \\
Seeing FWHM				& $0\farcs8$ \\
Calibration Uncertainty		& 1.3\% \\
Astrometric Uncertainty 		& $0\farcs03$--$0\farcs1$ \\
90\% Completeness Limit 	& 1.38 $\mu$Jy (23.55 mag.)\\
Spurious Fraction			& $<3.2\%$ 
\enddata
\end{deluxetable}

\section{Colors in the WIRCam and IRAC Bands}
\label{sec_color}

We next used the WIRCam $K_S$ image and catalog as priors
to analyze the fluxes of galaxies in the GOODS-N \emph{Spitzer} 
IRAC images at 3.6--8.0~$\mu$m (M.\ Dickinson et al., 
in preparation).  Similar work has been done previously.  In both the
GOODS-N \citep{laidler07} and the GOODS-S \citep{grazian06}, ACS 
$z^\prime$-band images were used as priors to determine fluxes in the 
IRAC bands.  While our $K_S$ image has
a lower resolution than the ACS $z^\prime$ images, the waveband 
is much closer to the IRAC bands, making it a better choice for this
type of analysis.

The primary goal is to better measure the colors of objects between
the IRAC and $K_S$ bands.  The nearly $2\arcsec$ resolution of IRAC 
is significantly worse than the typically sub-arcsecond resolution 
in the optical and NIR, but it is not much larger than the sizes of 
galaxies. This makes color measurements extremely uncertain.  Normally 
one would need to sacrifice resolution in the optical and NIR bands 
to achieve this.  The approach we describe below does not have
this disadvantage.  A secondary goal is to better separate IRAC 
fluxes of objects that are close to each other.  There are nearly 
20,000 $K_S$ detected source in the GOODS-N IRAC area, but SExtractor 
can only directly detect $11,000$ sources 
in the 3.6~$\mu$m IRAC image.  One reason for the smaller number of 
sources in the IRAC image is that many objects are blended under the 
IRAC resolution.  We will demonstrate that our method provides 
reasonable estimates of the IRAC fluxes of lightly blended objects.

\subsection{Methodology}

The basic idea is to convolve the $K_S$ image of each galaxy 
individually with a scaled convolution kernel to construct a model 
image for an IRAC image.  The scaling of the kernel varies between
galaxies because galaxies have different 
IRAC-to-$K_S$ flux ratios (i.e., colors).
The set of scaling factors that produce the best model image
are then the best estimates of the flux ratios.
 
It is immediately obvious that the results would be sensitive to the 
evaluation of whether a model best approximates the IRAC image, 
as well as whether the chosen convolution kernel well represents the 
PSF.  In \citet{laidler07} and \citet{grazian06}, 
the model that produced the minimum $\chi^2$ was considered as the best 
model.  In such analyses, $\chi^2$ is very sensitive to the mismatch 
of the PSF, and unfortunately, there is a huge uncertainty in the 
PSFs in the IRAC bands.  This was pointed out by 
Grazian et al.\ and is also clearly shown in Figure~\ref{fig_subtraction}.  
It is thus unclear to us whether or not $\chi^2$ analyses are the best 
approach.  Here we present an alternative approach based on the widely 
used image deconvolution technique in radio astronomy---CLEAN 
\citep{hogbom74}.  Instead of minimizing the $\chi^2$ of the pixels, 
CLEAN determines the scaling of the kernels based on the fluxes of 
the image components.  This is very attractive to us since our true 
goal is to be able to account for the fluxes detected in the IRAC 
images.  As long as the fluxes 
in an IRAC image can be cleanly removed by subtracting 
a model, we consider the model to be the best.  Below we describe 
the basic steps in our analysis of the galaxy colors.
We hereafter refer to our technique as REALCLEAN, since it is done 
in real space instead of Fourier space.

\subsubsection{PSF Construction}

The GOODS-N IRAC images have pixel scales of $0\farcs6$, so we
background subtracted and resampled them to the WIRCam $0\farcs3$ 
pixel scale.  During the resampling, the IRAC images were also 
slightly warped such that the positions of objects precisely match 
those in the WIRCam image\footnotemark[10].  To construct PSFs 
for the IRAC and WIRCam images, we selected $\sim20$ bright but 
unsaturated, point-like, and isolated objects in these images.  
The images of these objects were aligned, stacked, and normalized 
to form the PSFs.  The IRAC PSFs were then deconvolved by the WIRCam 
PSF with a maximum likelihood deconvolution of a few iterations.
The deconvolved IRAC PSFs were used as the convolution kernels to 
convolve the WIRCam image.  One may worry about the uncertainty 
in the deconvolution of the PSFs.  We also tried an alternative 
approach that does not require the deconvolution of the IRAC PSF, 
in which we measured colors in the IRAC images that are convolved 
with the WIRCam PSF and in the $K_S$ image that are convolved with 
the IRAC PSFs (see \S~\ref{xconv}). We did 
not see a systematic difference between the two approaches.  We thus 
adopted the deconvolved IRAC PSFs as the convolution kernels.

\footnotetext[10]{This is achieved with subroutines provided in the 
SIMPLE package.}

\subsubsection{Object Identification}

In the background-subtracted WIRCam $K_S$ image, for each detected object we 
identified pixels associated with that object.  This was done with the ``segmentation map''
provided by SExtractor.  This allowed us to apply different scaling factors to the
convolution kernels on different objects.  A minor drawback here is that very faint 
pixels in the outer parts of galaxies would be missed.  However, as long as these 
faint pixels do not have dramatically different colors from the brighter cores
of the galaxies, our measured colors can still be considered as good estimates of
the flux-weighted mean colors of galaxies.  Another minor drawback is pixels that 
contain emission from more than one object.  This would only become problematic 
when objects are not well resolved in the WIRCam image.  Since this is the 
fundamental limit of the WIRCam observations, we did not attempt to resolve
this issue.

\subsubsection{REALCLEANing Bright Objects}

To obtain model images that best approximate the observed IRAC images, we
use our REALCLEAN method, which
is especially useful when determining fluxes of blended galaxies.
In a REALCLEAN cycle, we first identify the brightest galaxy in the IRAC image.
Then we remove a small fraction (25\% at 3.6~$\mu$m down to 10\% 
at 8.0~$\mu$m; this factor is often called the ``gain'') of its 
flux from the IRAC image by subtracting a model image constructed by convolving 
the WIRCam image with the above mentioned kernel.  The flux is determined by 
placing a circular aperture at the object.  We used aperture diameters of $4\arcsec$, 
$4\arcsec$, $5\arcsec$, and $6\arcsec$ for the IRAC 3.6, 4.5, 5.8 and 
8.0~$\mu$m bands, respectively. Despite the fact that such aperture photometry 
is affected by nearby objects, because we only look at the brightest object
and because the gain is $\ll1$, we guarantee that we do not over-subtract the
object.  To speed up the REALCLEAN process, we subtract fractional fluxes 
from many objects in one REALCLEAN cycle.  However, in each REALCLEAN 
cycle we only subtract objects
that are sufficiently far away from each other that locally each subtracted object
is the brightest.  In the subsequent REALCLEAN cycle we look for the next 
brightest objects in the subtracted image and subtract small fractions of 
fluxes from them.  We repeat this iterative procedure until no objects in the 
residual IRAC image have fluxes brighter than a few $\sigma$.  

It is interesting to note that the above procedure is similar to the CLEAN 
deconvolution of radio interferometric data (see a summary in \citealt{cornwell99}) 
in many ways.  To limit the degrees of freedom, CLEAN is often limited 
to CLEAN windows or boxes within which emission is known or guessed to
be present. CLEAN also sometimes uses priors to improve
the speed of convergence of the solution.  These are exactly what our procedures
do: we only subtract IRAC fluxes at the locations of $K_S$ detected galaxies, 
and we use the WIRCam $K_S$ images of the galaxies as the model.

\subsubsection{Final Residual Flux Minimization}

Once REALCLEAN reaches a few $\sigma$, all objects have similar brightness and
REALCLEAN is no longer efficient and necessary.  Thus, we measured the residual IRAC 
fluxes of all of the WIRCam objects and subtracted them from the IRAC image all at once. 
This global subtraction unavoidably over-subtracts blended objects, especially those 
whose separations are less than the flux aperture sizes.  Since the global 
subtraction is carried out only after the objects are REALCLEANed to a few $\sigma$, 
the amount of over-subtraction is small.  Nevertheless, the over-subtraction
can be further minimized.  

We estimated the over-subtraction by measuring the
fluxes again in the residual image.  Then we re-iterated the global subtraction
by adjusting the subtracted fluxes and again measured the over-subtraction in the next 
residual image.  In each iteration, the adjustment of fluxes were small fractions (equal to 
the REALCLEAN gains) of the estimated over-subtraction.  This slowly brought the residual fluxes 
of galaxies in the subtracted image from negative (after the first global subtraction following the 
REALCLEAN cycles) to nearly zero after tens of iterations.  In the iterations, we closely 
watched the residual fluxes of galaxies to see if they converged to zero.  When the
residual fluxes no longer converged (typically after 40--60 iterations), we stopped the 
iterations.  Fluxes subtracted in the REALCLEAN stage and in this minimization stage
were added to make the total IRAC fluxes of the WIRCam $K_S$ detected galaxies.

For each IRAC band, there are two epochs of observations covering different areas
in the GOODS-N but with some overlap.  All the procedures described above, 
including the PSF construction, were carried out independently on each single-epoch 
image.  This is because the two epochs have different orientations of pixels and PSFs. 
For sources in the overlap region, we adopted the error weighted fluxes from the two 
epochs.

\subsubsection{Flux and Error Measurements}
\label{sec_irac_flux}

As mentioned, we used $4\arcsec$--$6\arcsec$ diameter apertures to 
measure fluxes, or in
other words, to evaluate whether the model best removes fluxes of galaxies.
Such small aperture sizes nicely cover the cores of the IRAC PSFs and 
minimize influences from nearby sources.  We also derived aperture corrections from
the convolution kernels.  However, the proper correction is a function of the sizes
of the galaxies, which are not exactly point sources under the IRAC resolution.  This 
biases the measured fluxes, especially when the apertures are small.
Our small apertures also miss the information in the outer wings of the PSFs.
To account for these, we repeated the REALCLEAN procedures with 50\% larger apertures 
and compared the results with those from small-aperture REALCLEANs.  Such aperture 
sizes are comparable to the sizes of the first diffraction rings in the IRAC PSF and 
are sufficiently large for the majority of galaxies in the GOODS-N.
We applied a constant correction factor (2\%--6\%) to all the small-aperture fluxes, 
derived from the large-aperture REALCLEAN on high S/N and isolated sources.  

We estimated flux errors in a similar way to those for the WIRCam
fluxes described in \S~\ref{k_catalog}.  We convolved the residual IRAC 
images with circular uniform kernels of amplitude 1.0 and measured the rms fluctuation locally.
This way, the measured noise included not only photon noise, confusion
noise, and correlated noise, but also rough estimates of the uncertainties caused by
imperfect object removal in the entire REALCLEAN and flux minimization processes.
IRAC fluxes and errors measured in this way are included in the $K_S$ selected 
source catalog in this data release (Table~\ref{master_cat}).

\subsection{Quality and Performance}

\subsubsection{Residual Images}
\label{sec_residual}

\begin{figure*}[ht!]
\epsscale{1.0}
\plotone{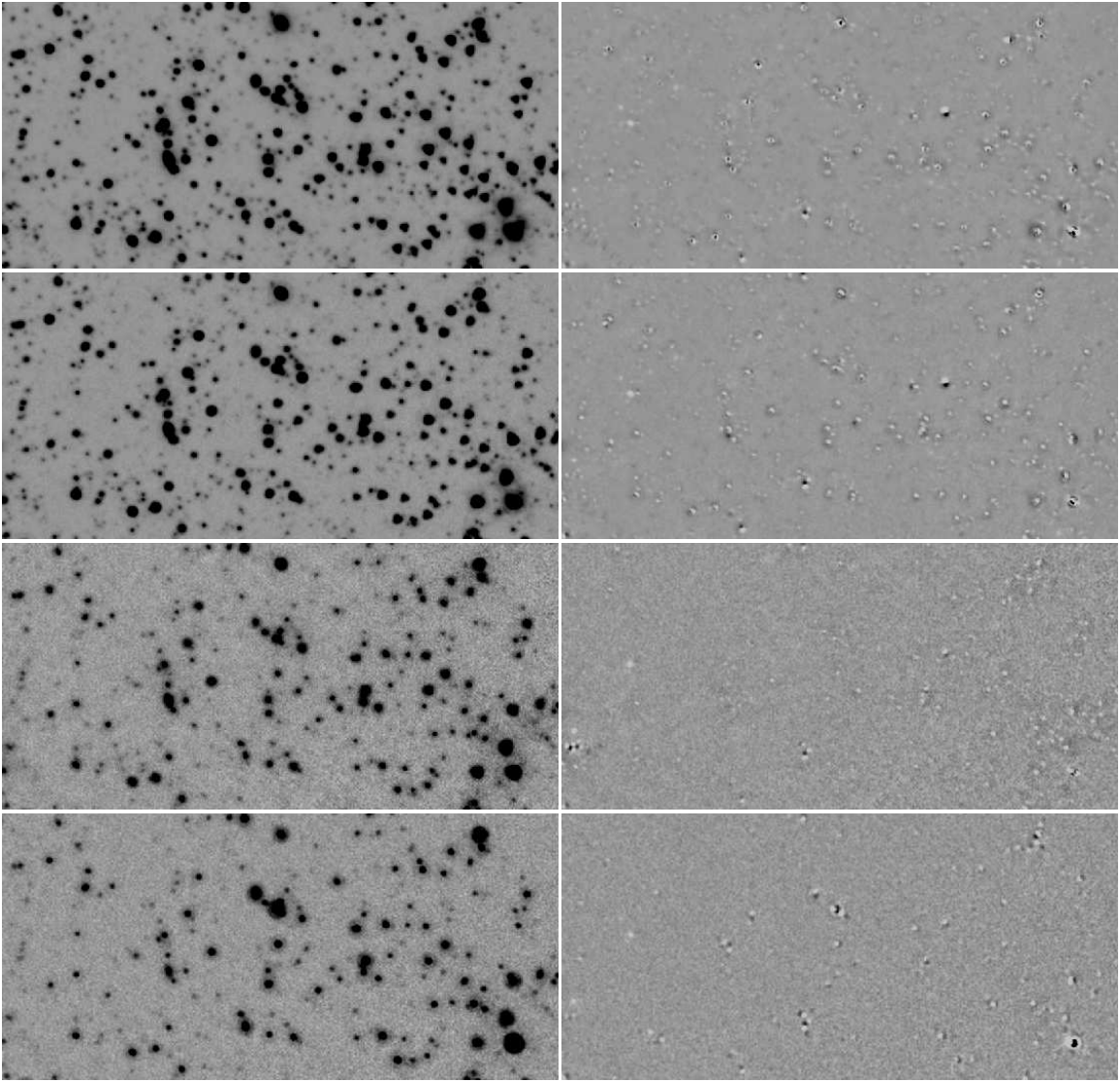}
\caption{Results of the REALCLEAN subtraction and residual flux minimization.
Images to the left are the original IRAC images in the HDF-N.  Images to
the right are the residual images after subtracting all sources with minimal
residual fluxes.  From top are the IRAC images at 3.6, 4.5, 5.8 and 8.0~$\mu$m.
Each residual image is shown with a brightness scale identical to its
original image. All images are shown with inverted scales.
\label{fig_subtraction}}
\end{figure*}

To evaluate the performance of the above REALCLEAN-based procedures, we 
first visually inspected the residual images after subtracting the best models. 
Figure~\ref{fig_subtraction} shows the HDF-N region of the IRAC images
before and after the subtraction.  Overall, the subtraction of IRAC objects
with our REALCLEAN method appears reasonably good in all IRAC channels.
However, minor residual effects can be seen.  There are faint halos around 
many objects in all channels after the subtraction, likely a consequence of the 
fact that our small flux apertures did not fully account for fluxes from slightly 
extended objects.  On the other hand, on compact objects there are residual 
patterns around the cores of the objects.  In the first two IRAC channels, this 
indicates a $\sim5\%$ uncertainty in the adopted IRAC PSF.  Because this 
occurs across the entire field, this is unlikely to be an effect of PSF variation or 
position-dependent mis-registration.  This is the most severe in the first two 
IRAC channels, suggesting that under-sampling of the PSF by the large IRAC 
pixels may have played a role.  However, the exact reason remains unclear to us.  
The 5.8~$\mu$m channel has the cleanest subtraction of IRAC objects.  
It gets slightly worse again in the 8.0~$\mu$m channel, likely because the 
8.0~$\mu$m morphology becomes sufficiently different than that in the 
$K_S$-band.

\begin{figure*}[ht!]
\epsscale{0.38}
\plotone{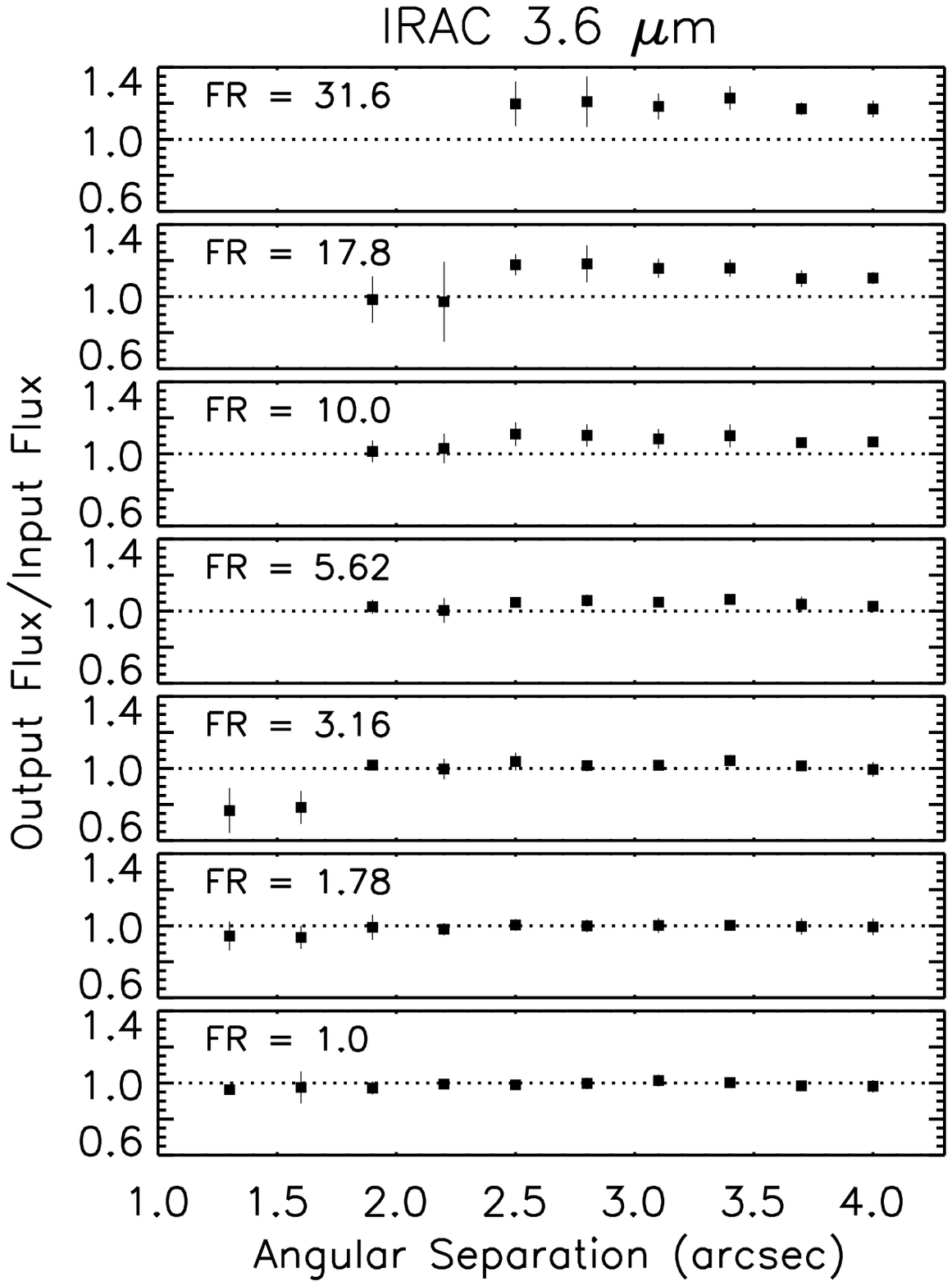}
\plotone{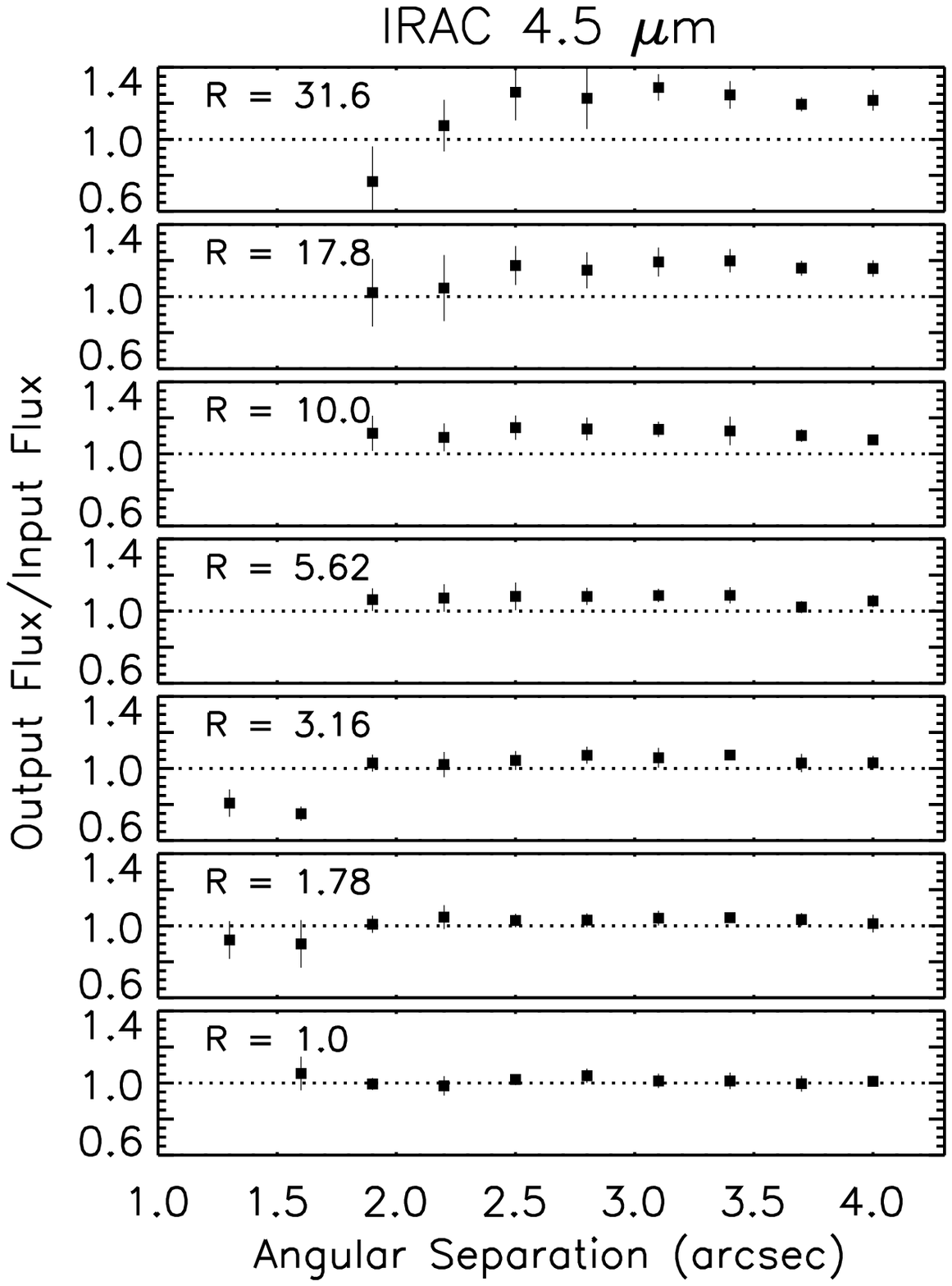}
\vskip0.5cm
\plotone{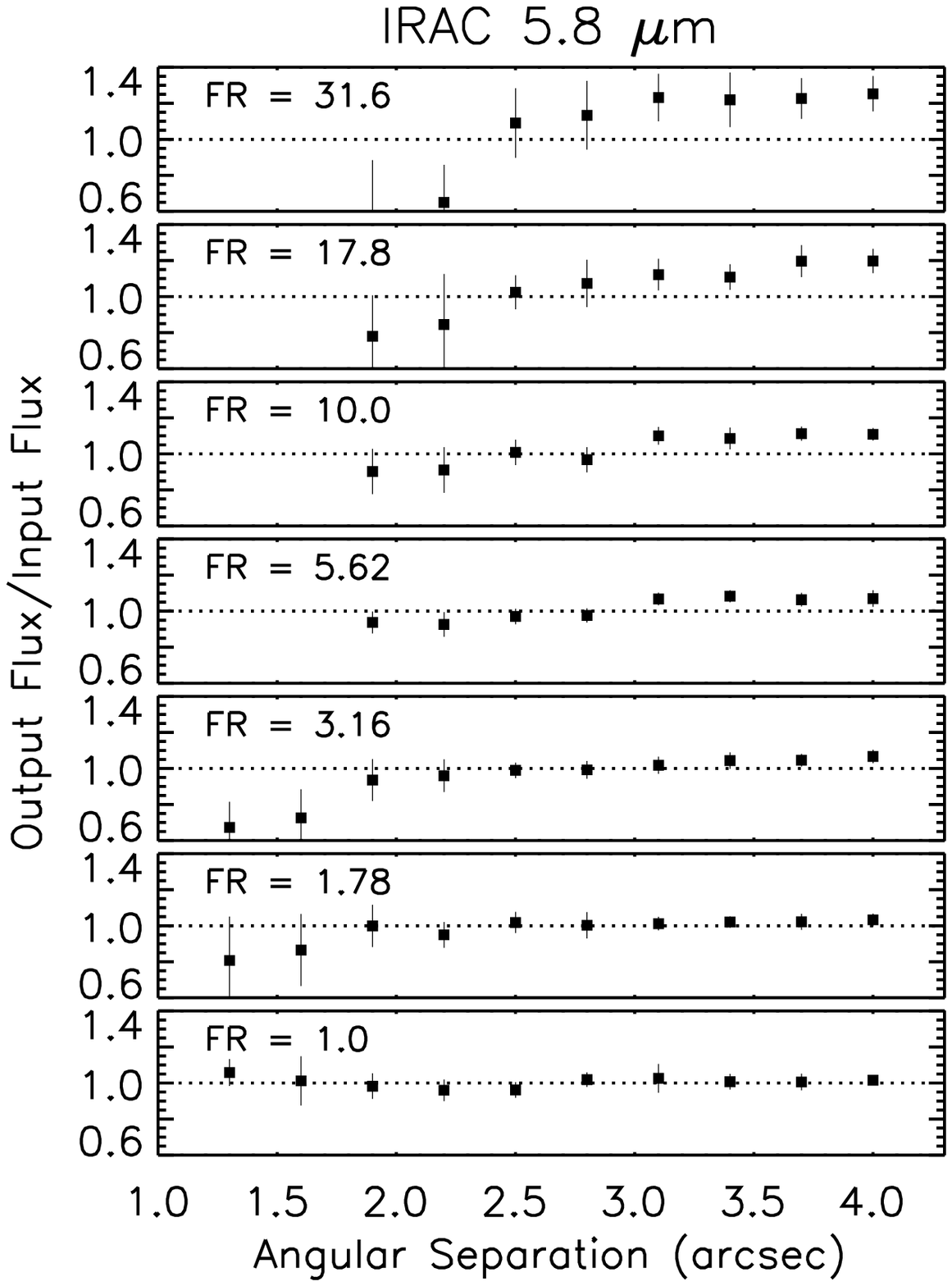}
\plotone{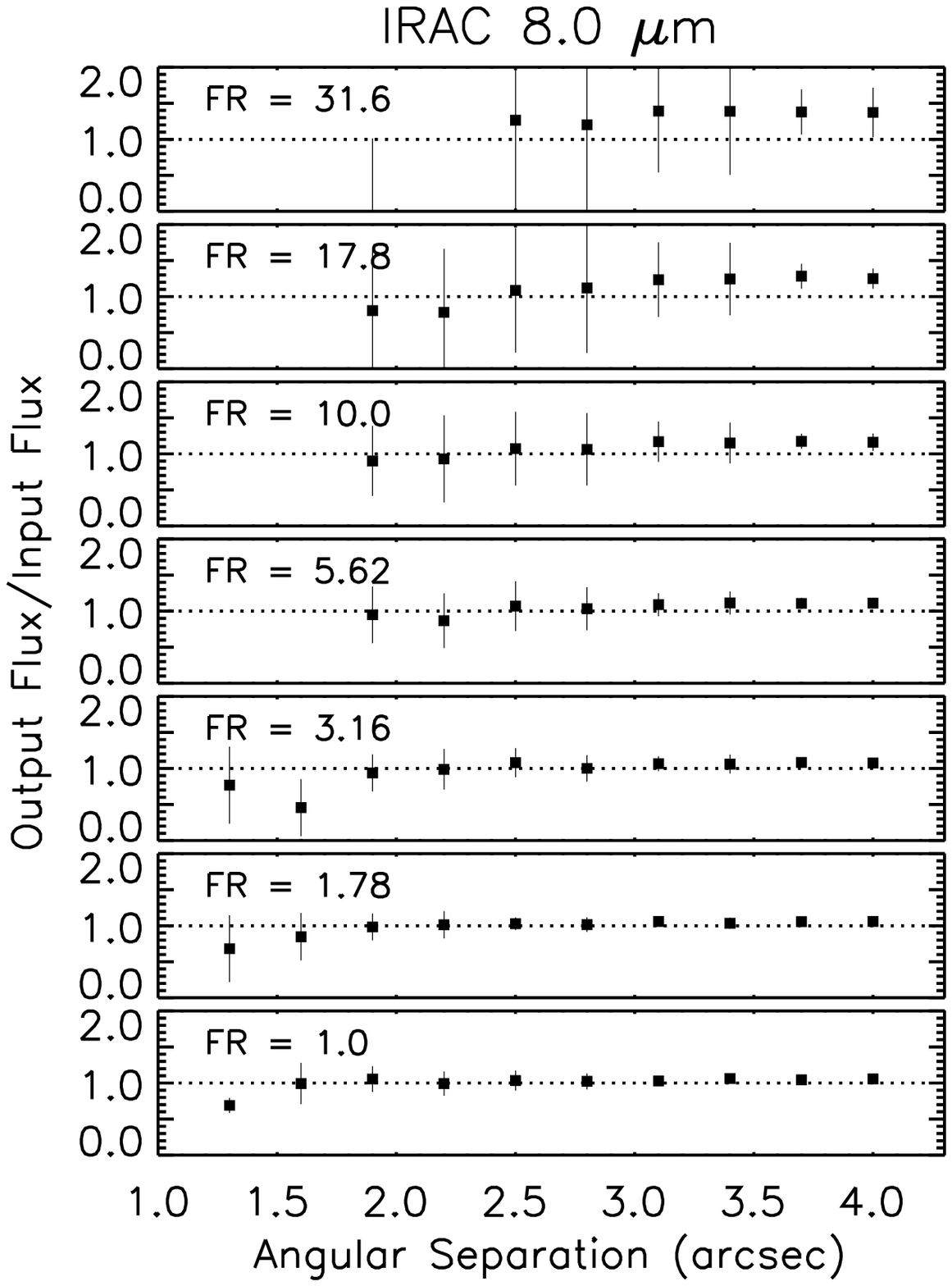}
\caption{Simple Monte Carlo simulations of REALCLEAN on two blended objects.
Four large panels show results from the four IRAC channels.  The simulations include
various flux ratios and angular separations for the blended pairs. Results of different flux
ratios (FR) are shown in the sub-panels.  Vertical axes show the output-to-input
flux ratios of the fainter objects between the blended pairs.  Error bars show the
dispersion among various realizations.  Note that the 8.0 $\mu$m panel has a different 
vertical scale. The lack of data points in some regions of the plots means that
even the SExtractor cannot separate the blended sources in the WIRCam images.
REALCLEAN cannot deal with such cases since it is based on the SExtractor WIRCam 
catalogs.
\label{fig_sim_clean}}
\end{figure*}

\subsubsection{Monte Carlo Simulations}

The above visual inspection of the residual images qualitatively shows that 
\emph{overall} our REALCLEAN approach can account for the
IRAC fluxes of $K_S$-band detected objects and that there are no 
catastrophic failures in subtracting off the 
IRAC objects. However, this does not guarantee that fluxes are
correctly assigned to objects when multiple objects are blended by the IRAC
PSF.  To test this, we carried out simple Monte Carlo simulations.

We simulated objects of high S/N ($\sim20$) blended by nearby objects.
The flux ratios between  the blending bright sources and the targets are
from 1.0 to 100.  The separations between the targets and the nearby sources 
are from $1\farcs3$ to $4\arcsec$, with various position angles.  
All objects are point-like and have flat spectra.  PSFs
derived from the two epochs of the real IRAC images and the real WIRCam image were 
used to create the associated simulated images.  All the images were processed
identically to the above REALCLEAN and catalog making procedures.

We shown the results of the simulations in Figure~\ref{fig_sim_clean}.
A few trends can be observed and are indeed as expected.
REALCLEAN performs the best when the fluxes ratios are $<10$ and separations are
$\gtrsim2\arcsec$.  At very small separations of $1\farcs3$, REALCLEAN can
also correctly recover the fluxes if the nearby objects are less than $2\times$ brighter
then the targets.  Among the $\sim15000$ real sources detected at 3.6 $\mu$m in our
final catalog (\S~\ref{sec_combined_catalog}), 90\% of their brightest neighbors
fall in the above ranges of flux ratio and angular separation.

There no data points in regions of large flux ratios and small separations
(including flux ratios of greater than 31.6, which are not shown).  
These are the cases in which even SExtractor cannot resolve the blended 
sources in the WIRCam images.  REALCLEAN cannot determine the IRAC 
fluxes if the sources are not resolved by SExtractor in the WIRCam images.  
At the boundary when the SExtractor can barely resolve the blended pairs, 
REALCLEAN either over-estimates (large separation) or under-estimates (small
separation) the fluxes of the targets.  The amount of over and under-estimate
can be up to 40\%.  Here we are pushing the limits of both REALCLEAN
and SExtractor.  In our catalog, the fractions of sources with $<4\arcsec$
neighbors that are $>17.8\times$ and $>31.6\times$ brighter are 5.7\% and
3.6\%, respectively.

The above results imply good REALCLEAN fluxes for the majority of sources in our catalog.
A small number of sources may have IRAC fluxes affected by very bright neighbors. 
The effect is generally not $\gg40\%$ before SExtractor becomes unable to detect 
the faint sources.  The caveat here is that the above simulations are very simple.  
One may conduct much larger simulations with variable object morphology, S/N level, and 
number of blended sources.  We do not think these are particularly useful given the 
large uncertainty in the IRAC PSF discussed in \S~\ref{sec_residual}.  Realistic simulations 
should take into account such uncertainties, folding in how the PSF is under-sampled by the
IRAC pixels and how the observations were carried out and reduced to better 
sample the images (e.g., dither and drizzle), which is out of the scope of this paper.  
Without good knowledge about the PSF, going further beyond the above simulations does not 
likely help on understanding the performance of REALCLEAN.  In what follows, we will rely 
on simple consistency checks within the available data, paying special 
attention to blended galaxies.

\subsubsection{Overlap Region Between Epochs 1 and 2}
\label{sec_overlap_region}

\begin{figure*}[ht!]
\epsscale{0.95}
\plotone{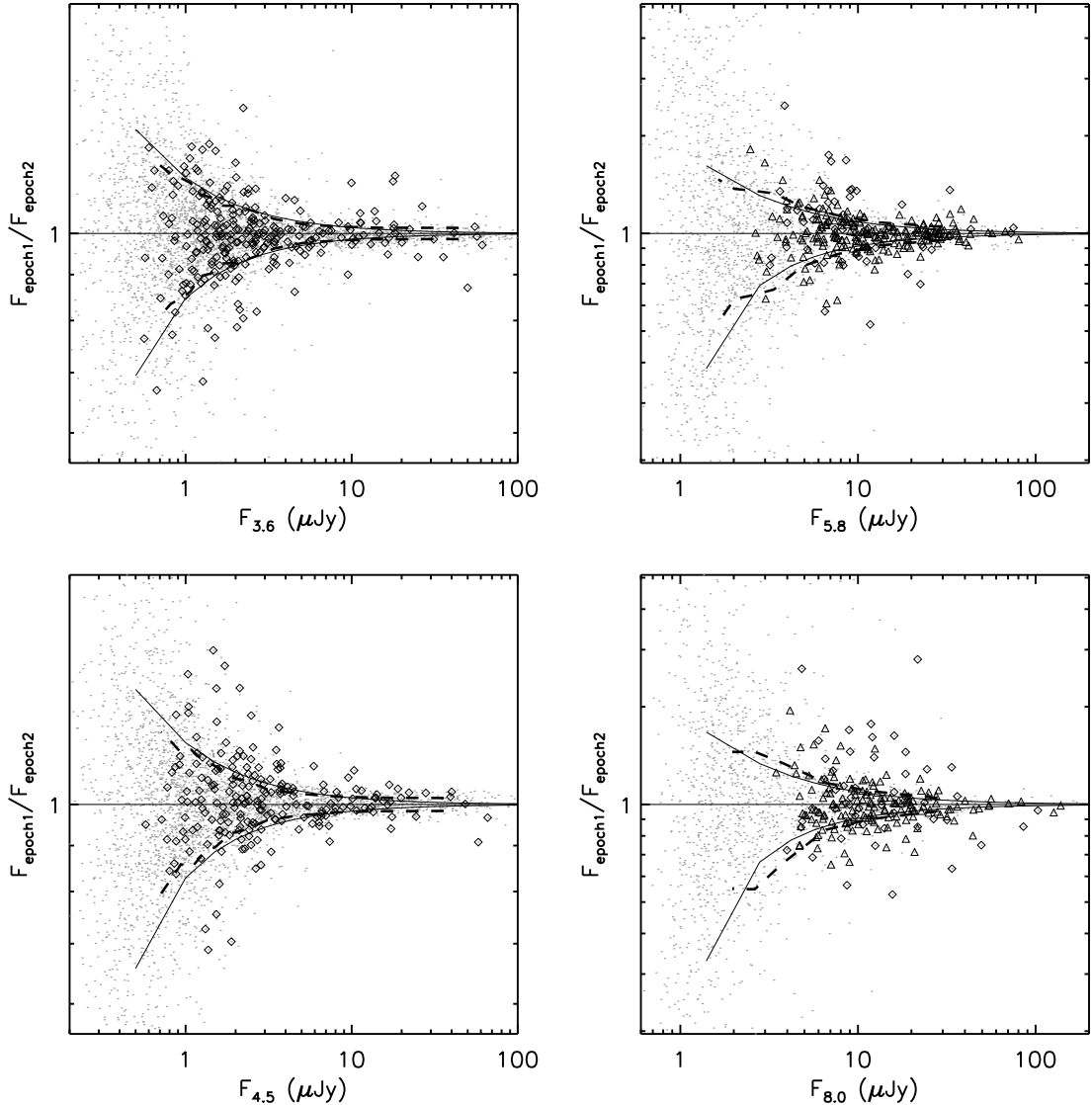}
\caption{Comparison of the IRAC fluxes derived from the two epochs.
\emph{Dots} show the full sample. \emph{Diamonds} and \emph{triangles} show 
galaxies whose nearest $K_S$ neighbors have angular separations of 
$1\arcsec$--$2\arcsec$ and $2\arcsec$--$3\arcsec$, respectively.  
\emph{Dashed curves} show the dispersion of the full-sample data. 
\emph{Solid curves} show the expected dispersion based on flux
uncertainties derived in \S~\ref{sec_irac_flux}.  The $y$-axes have logarithmic scales.
\label{fig_epoch1vs2}}
\end{figure*}

The GOODS-N IRAC observations consist of data from two epochs.
Both the PSFs and the pixels in these two epochs have different orientations.
The two epochs of imaging overlap in approximately 35~arcmin$^2$ of
area at the center of the GOODS-N.  Since we ran our REALCLEAN procedures
on the two epochs separately, we compared our results in the overlap
region to see if they are consistent.
Figure~\ref{fig_epoch1vs2} shows the ratios of the fluxes derived from epochs 1
and 2.  The \emph{dashed} curves in each plot show the dispersion of 
the flux ratios, and the \emph{solid} curves show the expected dispersion
based on flux errors estimated in \S~\ref{sec_irac_flux}.
The fact that the two curves are very close to each other is a strong
indication that our error estimate is realistic and that the uncertainty 
of the fluxes derived from our REALCLEAN procedure is limited by the noise
in the data.  

We examined the IRAC sources that have
nearby $K_S$-band detected objects with separations of $1\arcsec$--$2\arcsec$
(\emph{diamonds} and $2\arcsec$--$3\arcsec$ 
(\emph{triangles}).  In the $1\arcsec$--$2\arcsec$ category, the distributions 
of the flux ratios are not significantly different from the main sample, except that 
at 8.0~$\mu$m.  At 8.0~$\mu$m the $2\arcsec$--$3\arcsec$ category 
has a distribution similar to the main sample.  These results suggest that
our procedures produce consistent results against slight changes in the PSF
for blended sources with separations greater than $1\arcsec$ at 3.6, 4.5, and
5.8~$\mu$m and $2\arcsec$ at 8.0~$\mu$m.  In the $K_S$ catalog for
sources with S/N $>5$, the median separation between galaxies and their
nearest neighbors is $4\farcs1$, and 99.5\% of sources do not have a nearest
neighbor within $1\arcsec$.  This suggests that nearly all of the fluxes
we derived for the first three IRAC channels are reliable.  On the other hand,
among the $\sim5000$ 8.0~$\mu$m detected sources, approximately 10\%
of the sources have nearest neighbors at $<2\arcsec$. Their fluxes are
more problematic, if Figure~\ref{fig_epoch1vs2} is indicative.  However, the
percentage would be lower if we only considered sources whose nearest 
neighbors were significantly brighter than themselves.  One can imagine that
if sources with similar fluxes are blended, REALCLEAN naturally 
assigns similar fluxes to both if the gain is sufficiently small (which is the
case for our 8.0~$\mu$m REALCLEAN).  On the other hand, if sources with
different fluxes are blended, the procedure might fail, especially on the fainter 
member of the pair.  These are also indicated in our Monte Carlo simulations.

\begin{figure*}[ht!]
\epsscale{1.0}
\plotone{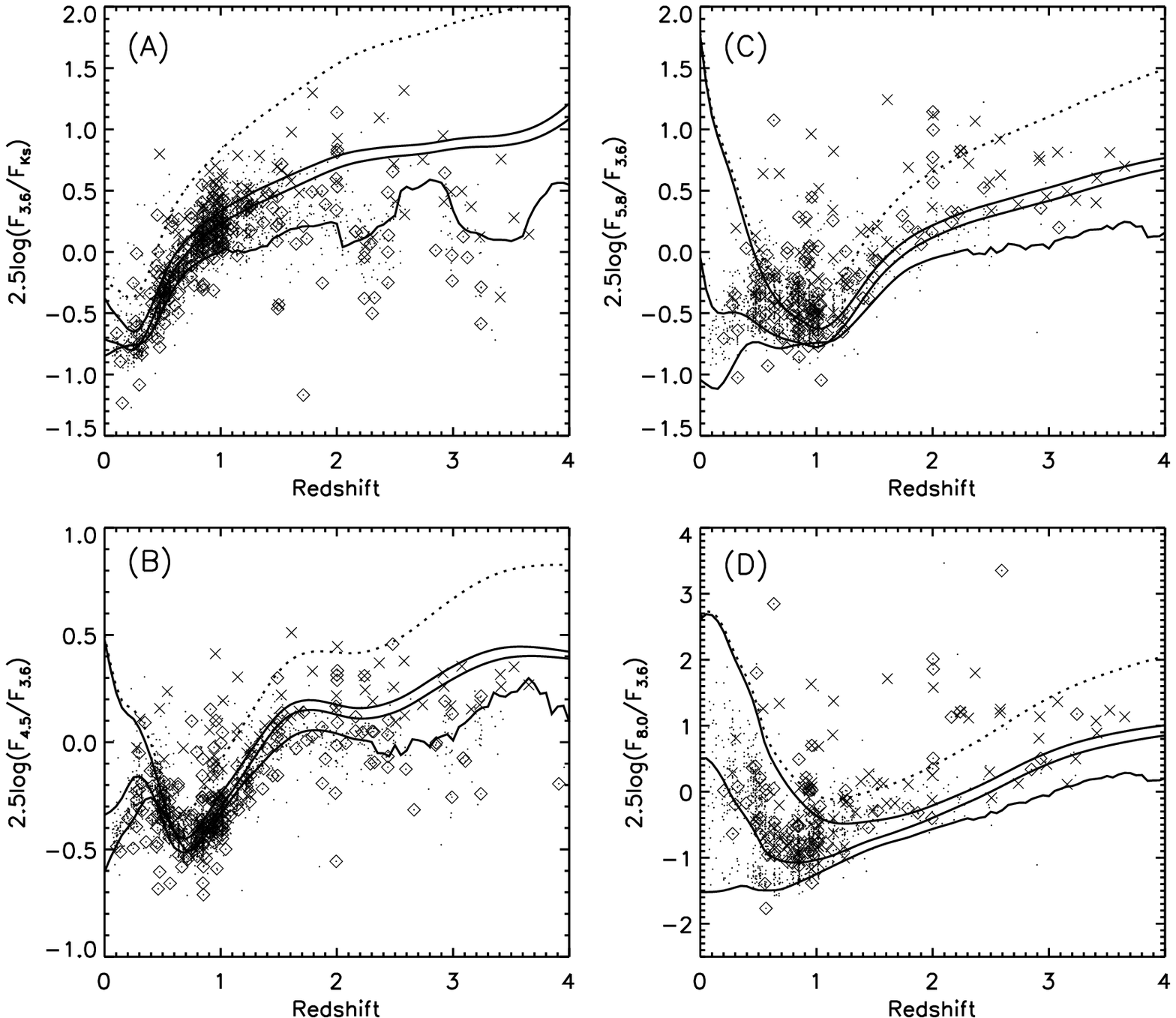}
\caption{IRAC and $K_S$-band colors vs.\ redshift for spectroscopically
identified galaxies.  Only galaxies with S/N~$>10$ at each waveband are plotted (\emph{dots}).
\emph{Diamonds} show galaxies whose nearest neighbors are within $1\arcsec$--$2\arcsec$. 
\emph{X-marks} show X-ray selected AGNs (hard or soft X-ray luminosities 
greater than $10^{42}$ erg s$^{-1}$).   
\emph{Solid curves} show unreddened galaxy templates of, from red to blue,
M~82 (a dusty starburst), M~100 (a spiral), and a dustless starburst.  \emph{Dotted curves}
show the M~82 template with $A_V=2.0$.  
\label{fig_z_color1}}
\end{figure*}

\subsubsection{Comparison with SED Models}
\label{sec_compare_sed}

The key goal of our analyses is to obtain accurate colors for the WIRCam 
and IRAC sources.  We selected sources that have redshifts in the 
spectroscopic survey of B08 and compared their colors with 
model SEDs.  Here we used two SEDs with dust and polycyclic aromatic
hydrocarbon (PAH) features from \citet{silva98}, M~82 (a dusty starburst) and 
M~100 (a spiral), and a dustless starburst template from \citet{kinney96} that has
a blue stellar continuum and strong recombination lines.
We also reddened the M~82 SED with $A_V=2.0$ using the \citet{calzetti00} 
extinction law.  These four SEDs represent a broad range of galaxy colors.
Figure~\ref{fig_z_color1} shows the comparison between the data and
the model SEDs. As in Figure~\ref{fig_epoch1vs2}, galaxies with nearest
neighbors of $1\arcsec$--$2\arcsec$ separations are shown with \emph{diamonds}.

Overall the derived IRAC and $K_S$ colors match the SED models 
excellently.
At $z\sim2.2$ and 3.4 in the $F_{3.6}/F_{Ks}$ ratio and $z\sim3$ in 
the $F_{4.5}/F_{3.6}$ ratio, there are substantial numbers of observed 
colors significantly bluer than the starburst model.  This is true
for both isolated galaxies and galaxies with close neighbors.  
We believe these blue ``outliers'' are not
artifacts of our REALCLEAN procedures but rather objects with emission lines even 
stronger than those in the starburst template.  
The lines are likely H$\alpha$ and [\ion{S}{2}] (entering $K_S$ at $z\sim2.2$),
H$\beta$ and [\ion{O}{3}] (entering $K_S$ at $z\sim3.4$), and [\ion{S}{3}] 
(entering the 3.6 $\mu$m band at $z\sim3$), all present in the starburst template.
We attempted to verify this with our spectroscopic data.
Unfortunately, we only have optical spectra on three of these galaxies, as most of the redshifts are adopted from
\citet{reddy06}.  The three sources have standard Lyman break galaxy spectra with strong 
\ion{C}{4} absorption.  In one case there is also Ly$\alpha$ emission.  In the other cases 
we do not cover Ly$\alpha$.  Galaxies with strong rest-frame optical 
(corresponding to $K_S$ and IRAC bands at these redshifts) emission lines have 
absorption lines in the rest-frame ultraviolet (corresponding to optical at these redshifts).
Given their rest-frame ultraviolet continua, these three galaxies are inevitably strong 
emission-line galaxies in the rest-frame optical.
This is consistent with strong emission lines causing the blue observed colors at $K_S$
and IRAC.  

There are  a few unexplained blue outliers at $z\sim1.4$--2 in $F_{3.6}/F_{Ks}$ and
at $z\sim2$ in $F_{4.5}/F_{3.6}$.  They might be 
fluxes mistakenly assigned by our REALCLEAN procedure.  However, if this
is the case, it would be very surprising if such outliers are not observed 
at all at $z<1.5$, where most of the spectroscopic samples lie.
An alternative explanation is that they are galaxies with 
misidentified redshifts. 
We have spectra for these galaxies, except for two whose redshifts were adopted from
\citet{reddy06}.  We inspected the available spectra and did not see obvious signs of 
misidentification.  Thus, the nature of these blue outliers remains unclear.
Nevertheless, we emphasize that the observed IRAC colors of the majority of 
the galaxies match excellently with the models at nearly all redshifts.  Combining 
this with our observations in \S~\ref{sec_overlap_region}, we are confident that
our REALCLEAN procedures correctly assigned fluxes to nearly all of the
IRAC sources.

\begin{figure}[ht!]
\epsscale{1.0}
\plotone{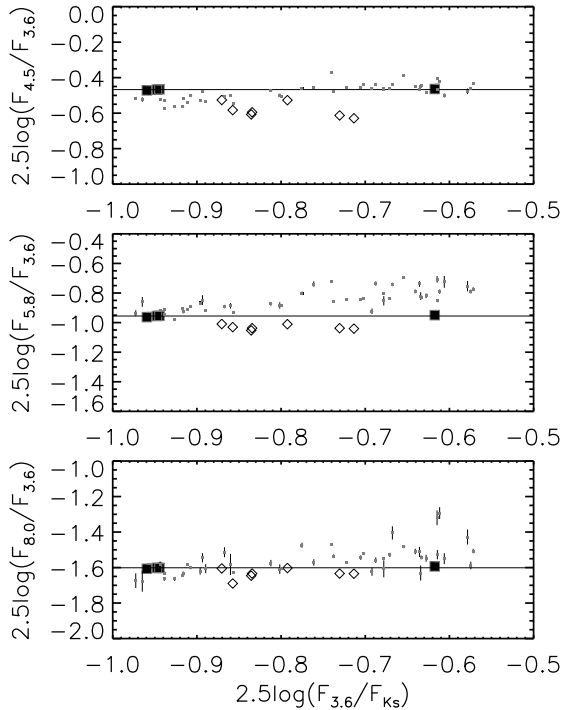}
\caption{IRAC colors of spectroscopically identified stars (\emph{dots}) and 
IRAC primary absolute calibrators.  A-type calibrators are shown as \emph{filled squares}
and K-type calibrators are shown as \emph{open diamonds}.  The mean colors of
the four A-type calibrators are shown with \emph{horizontal lines}.
\label{fig_star_color}}
\end{figure}

\subsubsection{Calibration}

The last performance test is a comparison with \emph{Spitzer} absolute
flux calibrators in \citet{reach05}.  Figure~\ref{fig_star_color} shows the
measured colors of the spectroscopically identified stars brighter than 
$K_S=19$ and the colors of the IRAC primary 
absolute calibrators in Reach et al.  Among the calibrators, A dwarfs are
shown with \emph{filled squares} and K giants are shown with \emph{open diamonds}.
Reach et al.\ found a displacement between the calibration derived from A and K stars,
especially in the first two IRAC channels.  The reason for this is unclear and the
authors decided to adopt an absolute calibration based on only A stars.
In Figure~\ref{fig_star_color} we see a gradual drift of the IRAC colors for our data
with respect to the IRAC calibrators, from blue to red in the $F_{3.6}/F_{Ks}$ color.  
However, since the IRAC absolute calibration is based on A stars, it is more
appropriate for us to compare A stars only.  At the very blue end where we find
three of the four A-type IRAC calibrators, our observed colors for the stars agree
excellently with the IRAC calibrators.  The largest offset is observed in the 
$F_{4.5}/F_{3.6}$ color (\emph{top} panel), where the stars in the left appear to be $\sim0.05$ 
magnitude bluer than the mean color of the A-type calibrators (measured from
stars with $x$-axis values $<-0.9$).  We thus 
conclude that our colors are well calibrated, with
systematic errors no larger than $\sim5\%$.

\subsubsection{Known Issues}
\label{issues}

Despite all the tests above, the IRAC fluxes derived by our REALCLEAN procedures
are not perfect.  The PSFs we used for REALCLEAN can only effectively remove fluxes
within $\sim4\times$ the FWHM.  There are about 10 or so very bright stars
with PSF wings, which are much broader than this.  Objects close to these
bright stars, especially those on the diffraction spikes and crosstalk features, have systematically
higher IRAC fluxes.  A more subtle effect is seen in a few very crowded 
regions where more than three objects heavily overlap.  In such configurations
objects that are close to the centers of the groups of objects can have systematically
higher fluxes.  Visual inspections identify only a handful of such examples in
each IRAC band.

\begin{deluxetable*}{lrrrrrrrrrrrr}
\footnotesize
\tablewidth{0pt}
\tablecaption{$K_S$ Selected Multi-Band Catalog \label{master_cat}}
\tablehead{\colhead{ID} & 
\colhead{R.A.}  & 
\colhead{Dec.}  &
\colhead{$F_{Ks}$}  & 
\colhead{$\sigma_{Ks}$}  & 
\colhead{$F_{3.6}$}  & 
\colhead{$\sigma_{3.6}$}  & 
\colhead{$F_{4.5}$}  & 
\colhead{$\sigma_{4.5}$}  & 
\colhead{$F_{5.8}$}  & 
\colhead{$\sigma_{5.8}$}  & 
\colhead{$F_{8.0}$}  & 
\colhead{$\sigma_{8.0}$} \\
\cline{4-13}
& \colhead{(J2000.0)} &
\colhead{(J2000.0)} & \multicolumn{10}{c}{($\mu$Jy)} }
\startdata
 54739 & 189.104027 & 62.278061 &  6032.930 &     0.229 &  2741.866 &  0.936  & 1662.465  &    0.632 &  1160.937   &   1.426  &  611.762    &  0.790 \\
 54740 & 189.054105 & 62.283076 &    94.032  &    0.235  &   82.640   &   0.158  &   53.823  &    0.203   &  56.101  &    0.860  &   38.938   &   0.841 \\
 54741 & 188.919686 & 62.287091 &     0.286   &   0.066   &   0.000   &   0.000  &    0.000   &   0.000   &   0.000  &    0.000  &    0.000   &   0.000 \\
 54742 & 189.482580 & 62.282344 &    54.345  &    0.233  &   31.454  &    0.173 &    23.463  &    0.171  &   17.371   &   0.468  &    9.386    &  0.356 \\
 54743 & 189.139434 & 62.284211 &    35.129  &    0.165  &   21.971  &    0.136  &   16.208  &    0.214  &   13.119  &    0.402  &    6.451  &    0.718 \\
 54744 & 188.774407 & 62.286231 &     1.029   &   0.213  &    0.000   &   0.000   &   0.000  &    0.000   &   0.000  &    0.000   &   0.000   &   0.000 \\
 54745 & 189.178137 & 62.286029 &     2.824   &   0.173  &    2.823  &    0.110   &   2.159  &    0.130  &    2.149  &    0.361  &    2.311    &  0.507 \\
 54746 & 189.747513 & 62.286519 &     0.184   &   0.075  &    0.000   &   0.000  &    0.000  &    0.000   &   0.000  &    0.000  &    0.000    &  0.000 \\
 54747 & 188.729697 & 62.286591 &     0.335   &   0.142  &    0.000  &    0.000  &    0.000   &   0.000  &    0.000  &    0.000  &    0.000   &   0.000 \\
 54748 & 189.369137 & 62.286251 &     2.069   &   0.103  &    2.484  &    0.121  &    1.975 &     0.133   &   1.891  &    0.432  &    1.085   &   0.262 \\
 54749 & 189.193693 & 62.286347 &     0.830   &   0.117  &   0.985   &   0.043  &    0.762    &  0.051  &    0.000   &   0.303  &    0.000    &  0.000 \\
 54750 & 189.071498 & 62.286398 &     1.692   &   0.110  &    2.196   &   0.094  &    1.674  &    0.117  &    0.894  &    0.353   &   0.752    &  0.254 
\enddata
\tablecomments{Table~\ref{master_cat} is published in its entirety in the electronic
edition of the \emph{Astrophysical Journal}.  A portion is shown here for
guidance regarding its form and content.  If the measured fluxes are less than the flux 
errors (S/N $<1$), zeros are assigned to the fluxes but the errors are given for upper limits.  Zero flux errors
mean that the sources do not have measurements (i.e., the $K_S$ sources are not in
the IRAC regions).}
\end{deluxetable*}

\subsection{Combined NIR and IRAC Catalog}
\label{sec_combined_catalog}

We combined the IRAC fluxes derived with the REALCLEAN procedures with the
$K_S$ source catalog to form the final multiband catalog.  The catalog is 
presented in Table~\ref{master_cat}.  The total number of 
sources is identical to the SExtractor $K_S$ catalog, but not every $K_S$
source contained a detection in the IRAC bands.  
As is the case for the $K_S$ fluxes, 
our IRAC flux errors are quite realistic (see \S~\ref{sec_overlap_region}
and Figure~\ref{fig_z_color1}), and we consider the IRAC sources with S/N~$<1$ 
to be undetected.  Table~\ref{tab_source_number} summarizes the number of 
detections and upper limits in the multiband catalog.

We have searched for missing sources in the REALCLEANed IRAC images.  
There may exist sources that are so red that they are detected in the IRAC
channels but not in the $K_S$ image.  These sources are mostly
too faint to be detected in the WIRCam $K_S$ image.  A small fraction of them
are blended with nearby sources in the $K_S$ image and not extracted by 
SExtractor. To search for such sources, we combined the REALCLEANed 
(those in the right hand side of Figure~\ref{fig_subtraction})
3.6 and 4.5 $\mu$m IRAC images and masked the $K_S$ detected objects.
This naturally excludes sources that are too close to bright objects.  We visually
inspected the REALCLEANed images and look for rermaining sources there.
We excluded sources that are likely cosmic rays in the IRAC
images, or other kinds of artifacts.  We also examined the 5.6 and 8.0 $\mu$m 
REALCLEANed images, and we found that all the sources seen in 
these two images can be extracted from the 3.6+4.5 $\mu$m REALCLEANed 
images.  Thus, the positions of such sources were measured in the 3.6+4.5 $\mu$m 
image and their fluxes were measured in all the individual IRAC images.  The $K_S$-band 
fluxes and flux errors are measured in the same way as described in \S~\ref{k_catalog}. 
The positions, $K_S$ fluxes, and IRAC fluxes of a total of 358 such sources are 
listed in Table~\ref{remain_cat}.  All of them have detections in at least two 
IRAC bands, and 112 of them are detected in all four bands.
As can be seen in Figure~\ref{fig_subtraction}, source extractions in the REALCLEANed 
images are difficult and can be severely affected by the imperfect source 
subtraction described in \S~\ref{sec_residual}.  We warn the
readers that Table~\ref{remain_cat} is very likely an \emph{incomplete} list of $K_S$-faint 
IRAC sources.

\begin{deluxetable}{lcccc}
\tablewidth{0pt}
\tablecaption{Number of Sources and Detection Limits in the Multi-Band Catalog \label{tab_source_number}}
\tablehead{\colhead{Waveband} & \colhead{Total Number} & \colhead{Detections} & 
\colhead{Upper Limits} & \colhead{Depth ($\mu$Jy)}}
\startdata
$K_S$ 		& 94951  	& 75187 	& 1784 	& 0.18\tablenotemark{a}  \\
3.6 $\mu$m 	& 16950	& 15018 	& 731   	& 0.11 \\
4.5 $\mu$m 	& 16631	& 14015	& 1000 	& 0.12 \\
5.8 $\mu$m 	& 16870	& 8481	& 4551 	& 0.42 \\
8.0 $\mu$m 	& 15184	& 7245	& 4118 	& 0.42 
\enddata
\tablenotetext{a}{If we limit the $K_S$ sample to the IRAC area, then the median 
1$\sigma$ error decreases to 0.12~$\mu$Jy.}
\tablecomments{The total number column gives the number of IRAC flux 
measurements at the locations of the $K_S$ objects.
They are mainly determined by the area coverage of the IRAC images. 
Detections are measurements with 
$>3\sigma$ fluxes.  Upper limits are measurements with $<1\sigma$ fluxes.  
Depths are median 1$\sigma$ errors among sources detected at 3$\sigma$ 
for the entire maps.}
\end{deluxetable}

\begin{deluxetable*}{rrrrrrrrrrrrr}
\tablewidth{0pt}
\tablecaption{IRAC Sources without $K_S$ Counterparts \label{remain_cat}}
\tablehead{\colhead{R.A.}  & 
\colhead{Dec.}  &
\colhead{$F_{Ks}$}  & 
\colhead{$\sigma_{Ks}$}  & 
\colhead{$F_{3.6}$}  & 
\colhead{$\sigma_{3.6}$}  & 
\colhead{$F_{4.5}$}  & 
\colhead{$\sigma_{4.5}$}  & 
\colhead{$F_{5.8}$}  & 
\colhead{$\sigma_{5.8}$}  & 
\colhead{$F_{8.0}$}  & 
\colhead{$\sigma_{8.0}$} \\
\cline{3-13}
\colhead{(J2000.0)} &
\colhead{(J2000.0)} & \multicolumn{10}{c}{($\mu$Jy)} }
\startdata
  189.049062  & 62.133903    & 0.227    & 0.091   &  0.945   &  0.128   &  0.894  &   0.160  &   1.597  &   0.625  &   1.049  &   0.385\\
  189.086443  & 62.134109    & 0.000    & 0.108   &  0.400   &  0.127   &  0.480  &   0.151  &   0.000  &   0.247  &   0.000  &   0.614\\
  189.049452  & 62.138386    & 0.000    & 0.074   &  0.377   &  0.078   &  0.110  &   0.080  &   1.870  &   0.560  &   0.741  &   0.377\\
  189.209527  & 62.140676    & 0.000    & 0.114   &  0.908   &  0.181   &  1.037  &   0.191  &   1.419  &   0.731  &   1.270  &   0.541\\
  189.215522  & 62.144264    & 0.000    & 0.094   &  0.638   &  0.212   &  1.049  &   0.190  &   0.000  &   0.284  &   1.507  &   0.590\\
  189.241636  & 62.145789    & 0.250    & 0.082   &  0.423   &  0.169   &  0.589  &   0.118  &   0.590  &   0.402  &   0.000  &   0.282\\
  189.246272  & 62.146429    & 0.217    & 0.105   &  0.468   &  0.151   &  0.623  &   0.122  &   0.377  &   0.324  &   0.000  &   0.258\\
  189.087076  & 62.150220    & 0.000    & 0.087   &  0.756   &  0.232   &  0.366  &   0.211  &   0.696  &   0.300  &   0.000  &   0.425\\
  188.998232  & 62.150675    & 0.000    & 0.095   &  0.269   &  0.117   &  0.581  &   0.159  &   0.000  &   0.337  &   0.426  &   0.335\\
  189.018589  & 62.153052    & 0.192    & 0.101   &  0.345   &  0.112   &  0.277  &   0.129  &   0.000  &   0.496  &   0.769  &   0.557\\
  189.045539  & 62.154243    & 0.000    & 0.093   &  0.572   &  0.129   &  0.511  &   0.146  &   1.249  &   0.549  &   0.000  &   0.438\\
  188.981101  & 62.154402    & 0.254    & 0.081   &  1.280   &  0.198   &  2.046  &   0.210  &   3.184  &   0.922  &   2.433  &   1.231
\enddata
\tablecomments{Table~\ref{remain_cat} is published in its entirety in the electronic
edition of the \emph{Astrophysical Journal}.  A portion is shown here for
guidance regarding its form and content.  If 
the measured fluxes are less than the flux errors (S/N $<1$), then zeros are 
assigned to the fluxes, but the errors are given for upper limits.}
\end{deluxetable*}

\begin{figure*}[ht!]
\epsscale{1.0}
\plotone{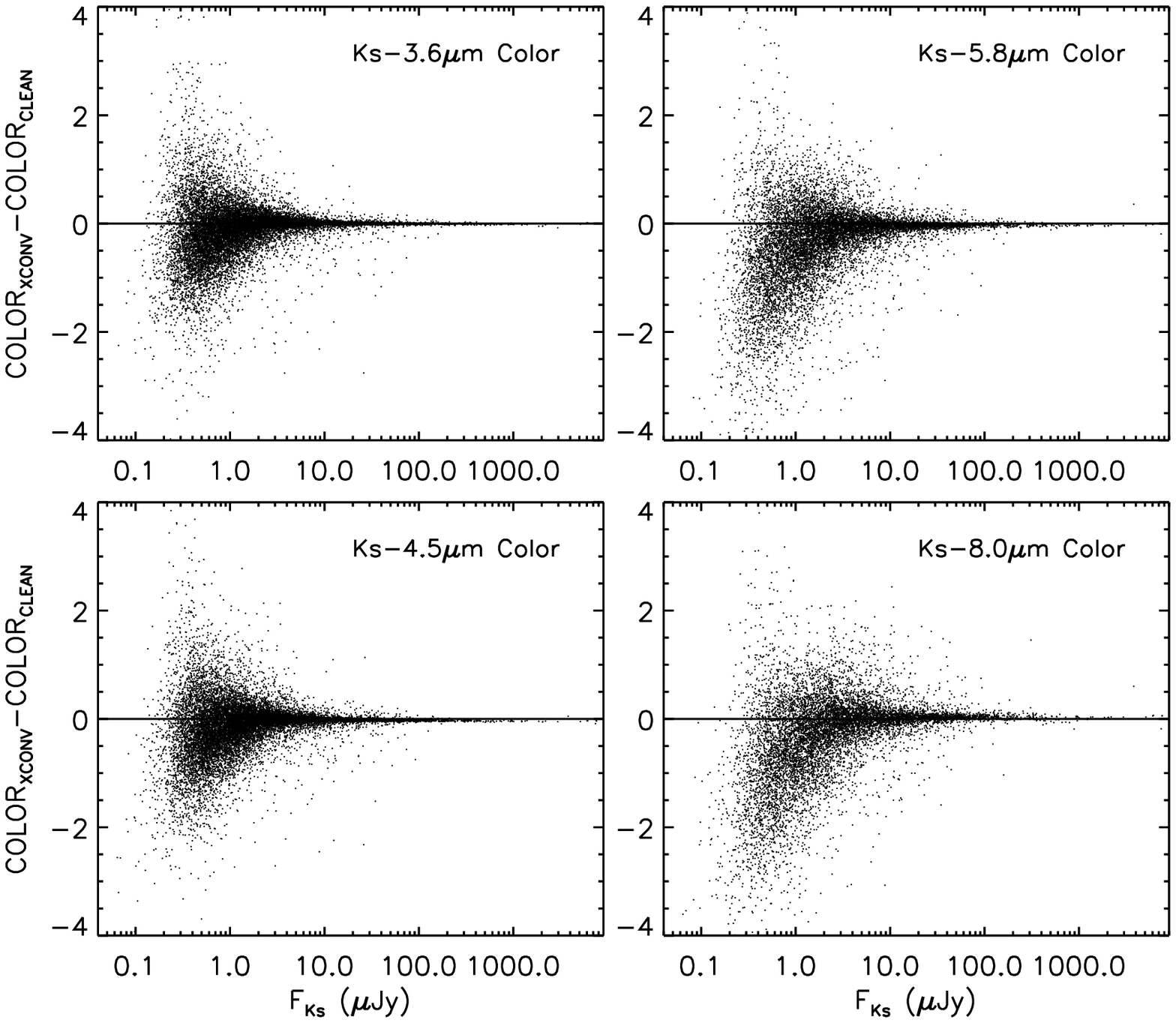}
\caption{Comparison between the colors derived with the XCONV and REALCLEAN 
methods.  All sources in both catalogs are plotted.
\label{fig_compare_colors}}
\end{figure*}

\begin{deluxetable}{rrrrr}
\tablewidth{0pt}
\tablecaption{$K_S$-to-IRAC Colors Derived with Cross-Convolution\label{xconv_cat}}
\tablehead{\colhead{ID}  & 
\colhead{$K_S-3.6$$\mu$m}  & 
\colhead{$K_S-4.5$$\mu$m}  & 
\colhead{$K_S-5.8$$\mu$m}  & 
\colhead{$K_S-8.0$$\mu$m} }
\startdata
 54739 &   -0.839 &   -1.434  &  -1.807  &  -2.430 \\
 54740 &   -0.135 &   -0.628  &  -0.576  &  -0.884\\
 54741 &   99.000 &    99.000  &  99.000 &   99.000\\
 54742 &   -0.579  &  -0.928  &  -1.283   & -1.879\\
 54743 &   -0.517  &  -0.862  &  -1.095   & -1.537\\
 54744 &   99.000 &   99.000 &   99.000 &   99.000\\
 54745 &    0.154  &  -0.067  &   0.166   &  0.174\\
 54746 &   99.000 &   99.000  &  99.000  &  99.000\\
 54747 &   99.000 &   99.000  &  99.000  &  99.000\\
 54748 &    0.365  &   0.190  &  -0.273  &  -0.473\\
 54749 &   -0.659  &  -0.817 &   99.000 &   99.000\\
 54750 &    0.283  &  -0.006 &   -0.249  &  -0.316
\enddata
\tablecomments{Table~\ref{xconv_cat} is published in its entirety 
in the electronic edition of the \emph{Astrophysical Journal}.  A 
portion (identical to that for Table~\ref{master_cat}) is shown here 
for guidance regarding its form and content.  Source ID and the total 
number of sources are both identical to that in Table~\ref{master_cat}.
All colors are in the AB magnitude system. 
Sources with no measurements are assigned 99.0.}
\end{deluxetable}

\subsection{Alternative Color Measurements with Cross-Convolution}
\label{xconv}

In the previous subsections, we described our REALCLEAN method of obtaining 
robust color measurements, for overcoming the issues of the very different PSFs  
and the blending of sources.  Here we describe an alternative method that can 
better handle the former issue but is more vulnerable to the latter.  
We started with the background subtracted and registered images used for
the REALCLEAN procedure. For each 
single-epoch IRAC image, we convolved the WIRCam $K_S$ image with the IRAC
PSF and the IRAC image with the WIRCam PSF.  We refer to this procedure
as cross-convolution (XCONV). We then used the 2-image
mode of SExtractor to detect sources in the unconvolved WIRCam image
and to measure the WIRCam and IRAC fluxes in the XCONVed images at
the locations of the detected WIRCam sources. The PSFs in the XCONVed
images have FWHMs of slightly larger than $2\arcsec$. After experimenting
with various photometry apertures, we found that a $3\arcsec$ aperture gives the
best compromise for S/N, inclusion of most fluxes from a wide range of
source structure, and blending of sources. Here we focus on the $K_S$-to-IRAC colors
measured this way, and we do not attempt to measure total fluxes.

The advantage of the XCONV method is clear: the PSFs are ideally 
matched and deconvolution of the IRAC PSFs is not required (unlike in
the REALCLEAN method).  The disadvantage is the severely degraded 
$K_S$ image resolution,
which prevents the extraction of information on both blended sources and faint sources. 
Because of these, we consider the XCONV colors of relatively bright
and isolated sources to be more robust than the REALCLEAN colors. 
However, as shown in 
Figure~\ref{fig_compare_colors}, the XCONV colors and the REALCLEAN colors turn
out to be remarkably consistent. The systematic offsets in Figure~\ref{fig_compare_colors}
between these two colors for $\sim1800$ $F_{K_S}>10$~$\mu$Jy objects 
are 0.002, -0.024, 0.031, and 0.030
(from 3.6 to 8.0~$\mu$m).  We examined the small number of objects in the figure that 
have offsets of $>0.05$ at $F_{K_S}>10$~$\mu$Jy. They are either close to 
bright objects (a weakness of the XCONV method) or are in regions with multiple
intermediate brightness objects (a weakness of the REALCLEAN method, as mentioned in
\S~\ref{issues}). 

The excellent consistency between the XCONV colors and the REALCLEAN colors further 
shows that our REALCLEAN method is robust.  We thus consider the REALCLEAN catalog as 
the primary product of this work, and we recommend it for most applications.
On the other hand, since the XCONV colors have unique advantages in their own right,
we also provide the XCONV colors in Table~\ref{xconv_cat}. We remind the reader
that the XCONV colors are less reliable on faint or blended sources. Such colors 
should be adopted with caution.

\section{Properties of the $K_S$ and IRAC Catalog}
\label{sec_properties}

Here we provide general descriptions of the properties of our $K_S$ 
and IRAC REALCLEAN catalogs. In subsequent papers we will present 
scientific analyses on high-redshift galaxies based on this catalog 
and other multiwavelength data.

\begin{figure*}[ht!]
\epsscale{1.0}
\plotone{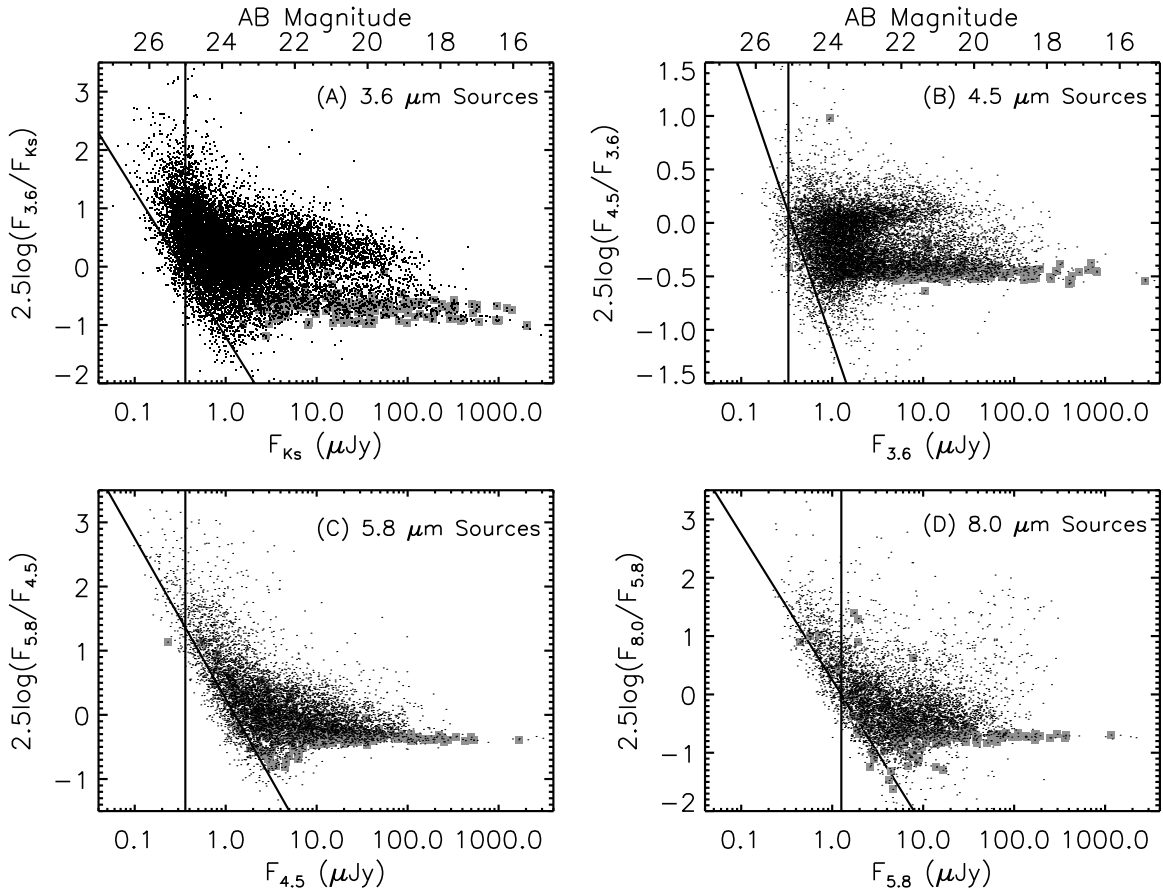}
\caption{$K_S$ and IRAC color-magnitude diagrams of IRAC selected sources.
\emph{Dots} show all sources selected at each corresponding IRAC band at $>3\sigma$.
\emph{Gray squares} show spectroscopically identified stars.
In each panel, \emph{solid lines} show the median $3\sigma$ detection limits of the 
two wavebands in consideration, adopted from Table~\ref{tab_source_number}.
\label{fig_color_mag-1}}
\end{figure*}

\subsection{Overlap with Other Catalogs}
\label{sec_cat_overlap}

The GOODS-N has extremely rich multiwavelength data and several 
catalogs have been published of sources in this region.  Here we 
summarize the overlap between these catalogs and our $K_S$ selected 
catalog.

In the GOODS-N and its flanking fields, the two most important optical 
observations are the \emph{HST} ACS imaging of the GOODS-N 
\citep{giavalisco04} and the Subaru SuprimeCam imaging of the 
$\sim0.45$~deg field around the GOODS-N \citep{capak04}.
In the 184~arcmin$^2$ \emph{HST} ACS region, there are 16326 $K_S$ 
selected sources with S/N $>3$ in our catalog and 39432 $z^\prime$ (F850LP) selected 
ACS sources in the GOODS v2.0 catalog.  With $0\farcs4$ search radii
($0.5\times K_S$ seeing FWHM), 14291 (87.5\%) $K_S$ sources have at least one 
ACS counterpart.  The number increases to 14883 (91.2\%) when the 
search radii are $0\farcs8$.  It increases much more slowly
beyond this. Among the 1443 optically faint $K_S$ sources, 
547 of them are detected at and 3.6~$\mu$m at $>3\sigma$.  
These are extremely unlikely spurious sources and appear to be truly 
optically faint NIR objects.  The rest of the 
optically faint $K_S$ sources are either outside the IRAC region (407 sources) or
do not have IRAC counterparts (489 sources).  The later case may be an 
indication of the existence of spurious sources in our $K_S$ catalog 
(see \S~\ref{sec_spurious}).  

We can also reverse the direction of the counterpart search.  Among
the 39432 ACS sources, 16982 (43.1\%) have at least one $K_S$ 
counterpart within $0\farcs8$.  The number decreases to 14853 (37.7\%) 
for $0\farcs4$ search radii.  These show not only that the $K_S$ depth 
is insufficient for detecting most of the ACS sources, but also that 
the $K_S$ resolution is too low for separating some close pairs of 
galaxies in the ACS image.

\citet{capak04} released a Subaru SuprimeCam catalog of the GOODS-N
region covering an area of 713~arcmin$^2$ with 48858 $R_C$+$z^\prime$ 
selected sources.  In the same region, there are 51649 $K_S$ selected 
sources with S/N $>3$, among which 73.2\% have counterparts in the Subaru catalog 
within $1\farcs2$.  The catalog of \citet{capak04} adopted a tighter 
S/N cut than our $K_S$ catalog.  This partially explains why the 
fraction with optical counterparts here is much lower than in the 
ACS case, despite the fact that the Subaru images are
very deep.  The large scattering halos around bright stars in the 
Subaru image create ``holes'' in the catalog, and $K_S$ sources do 
not have optical counterparts in such holes.

The X-ray \emph{Chandra} Deep Field-North (CDF-N) is fully covered 
by our $K_S$ imaging.  The \emph{Chandra} 2~Ms imaging \citep{alexander03}  
covers an area of approximately 420~arcmin$^2$ with an uneven sensitivity 
distribution.  There are 503 X-ray sources in the 2~Ms catalog, 
493 (98.0\%) of which have $K_S$ source counterparts within $2\farcs5$ 
search radii.  Among the 10 $K_S$-faint sources, four are close to 
the edge of relatively bright objects.  The brighter $K_S$ objects may have 
affected the detection of the X-ray counterparts at $K_S$, or themselves 
may be the correct counterparts.  To check whether the rest 6 are spurious 
X-ray sources, we look for their counterparts at other wavelengths.
Two $K_S$-faint X-ray source (source \#151 and \#183 in the catalog of
\citealp{alexander03}) have faint IRAC counterparts and are extremely
faint in the optical.  They are even close to two submillimeter sources detected at 1.1 mm
(AzGN10 amd AzGN12 in \citealp{perera08}).  Another (\#88) has a 
faint optical counterpart but is outside the IRAC region.  
These three are therefore genuinely $K_S$-faint X-ray sources.
Two $K_S$-faint X-ray sources (\#248 and \#371) are undetected in the Subaru optical, \emph{Spitzer}
3.6 to 24 $\mu$m, and VLA 1.4 GHz images. 
Even in the X-ray images they appear to be quite faint. 
They are more likely spurious X-rayt sources. Finally, one X-ray source 
(\#427) is not detected in any of the optical to radio images we have, 
but its detection appears quite convincing in the hard X-ray band.
Its nature is undetermined.

The \emph{Spitzer} GOODS Legacy Program also imaged the GOODS-N field with
Multi-Band Imaging Photometer for \emph{Spitzer} (MIPS, \citealp{rieke04}) 
at 24~$\mu$m.  The GOODS Data Release 1+ provided a 24~$\mu$m catalog
that is highly complete at 80~$\mu$Jy, containing 1199 sources.  With search radii of
$2\farcs5$, 1182 MIPS sources have counterparts in our $K_S$ catalog.  The rest
are all in low S/N regions in the MIPS images and appear to be spurious.  Thus,
our $K_S$ image seems to detect all $>80$~$\mu$Jy 24~$\mu$m sources, bearing 
in mind the possibility that some sources could be misidentified by the large search radii.

Finally, at the long wavelength side, there exist deep VLA 1.4 GHz imaging in the 
GOODS-N \citep{richards00}.  \citet{biggs06} re-reduced the VLA data and provided
improved images and catalog.  The catalog of Biggs \& Ivison contains 537 
$>30$~$\mu$Jy ($\sim5\sigma$) 1.4 GHz sources within a radius of $\sim24\arcmin$ 
centered at the HDF-N with a non-uniform sensitivity distribution.  A total of 450 radio
sources are within the $30\arcmin$ region of our $K_S$ image, 439 (97.5\%)
of which have $K_S$ counterparts within $1\farcs5$ of search radii.  The 11 
$K_S$-faint radio sources do not appear to be spurious in the radio image, although
some of them are very close to the $5\sigma$ limit at 1.4~GHz.  They are all in the outer 
shallower region of the $K_S$ image. We note that, in the outer part of the image, the
$K_S$ sensitivity decreases dramatically with exposure time (Figure~\ref{fig_layout}), 
while the radio sensitivity decreases slowly with the primary beam of the VLA.
We therefore conclude that the central part of our $K_S$ image is sufficiently deep for 
detecting nearly all of the 30~$\mu$Jy radio sources.

\begin{figure*}[ht!]
\epsscale{1.1}
\plotone{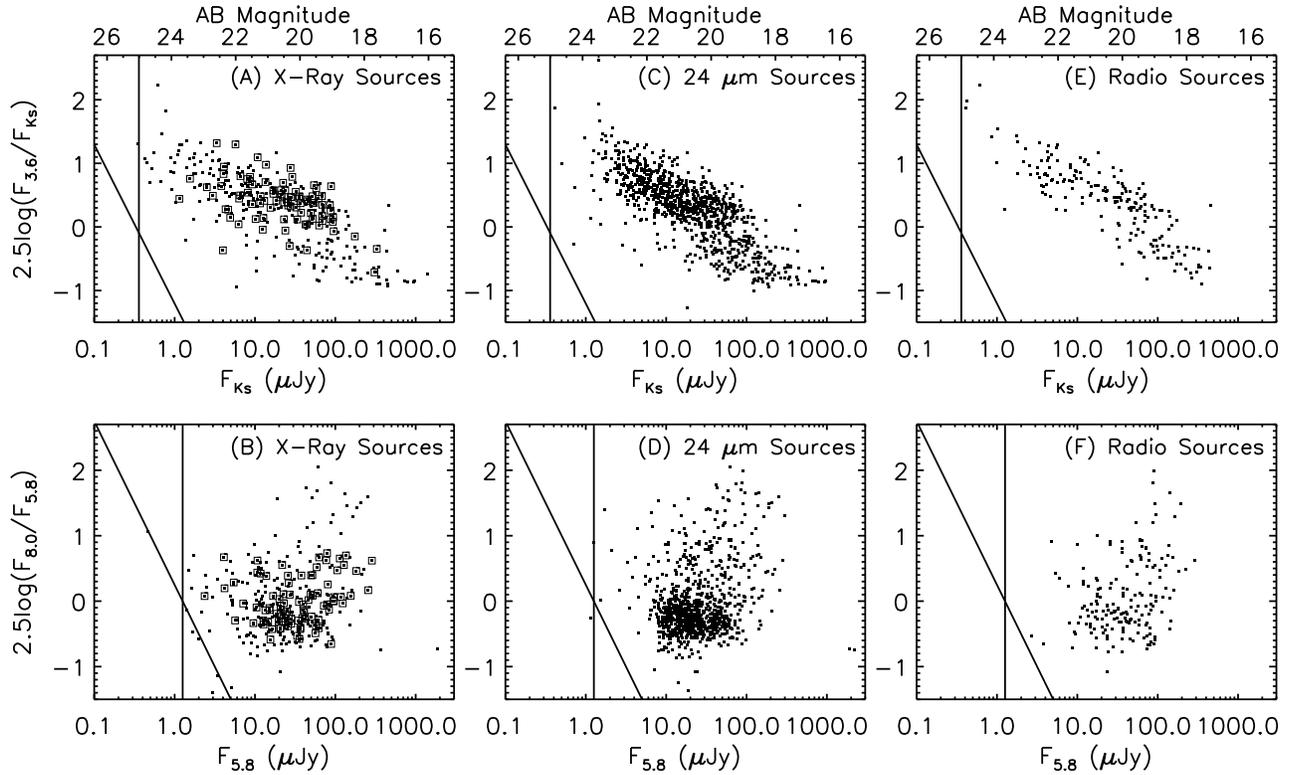}
\caption{$K_S$ and IRAC color-magnitude diagrams of X-ray, 24 $\mu$m,
and 1.4 GHz selected sources.  In panel (A) and (B), X-ray sources more luminous
than $10^{42}$~erg~s$^{-1}$ are enclosed by \emph{squares}.  \emph{Solid lines}
in each panel show median $3\sigma$ detection limits of the two wavebands in 
consideration.
\label{fig_color_mag-2}}
\end{figure*}

\subsection{Color-Magnitude Diagrams}

In Figure~\ref{fig_color_mag-1} we present color-magnitude 
diagrams for sources selected in each of the four IRAC bands.  The most 
apparent feature
is the color bimodality in the $F_{3.6}/F_{Ks}$ and $F_{4.5}/F_{3.6}$ colors.  
There are two distinct, nearly horizontal branches.  
The blue branches contain most
of the stars and low-redshift galaxies.  
The redshift divisions in $F_{3.6}/F_{Ks}$ and $F_{4.5}/F_{3.6}$ 
are $z\sim0.7$ and $z\sim1.2$, respectively.  At these two redshifts, galaxies 
quickly move across
the two branches, because of the rest-frame 1.6~$\mu$m SED bump in galaxies.  
This can also be seen in Figure~\ref{fig_z_color1}, which is based on our 
spectroscopic redshifts.  
The division at $z\sim1.2$ in the $F_{4.5}/F_{3.6}$ color is much more apparent
in Figure~\ref{fig_color_mag-1} because Figure~\ref{fig_z_color1} suffers a bit 
from the spectroscopic
redshift desert at $z\gtrsim1.4$.  The bimodality nearly disappears in the 
$F_{5.8}/F_{4.5}$ color and 
entirely disappears in the $F_{8.0}/F_{5.8}$ color.  
The main reason is that strong PAH emission from 
low-redshift starbursts significantly reddens the 
colors.  A minor reason is the relatively lower S/N at these longer IRAC bands.

In Figure~\ref{fig_color_mag-2} we show the color-magnitude diagrams for the
X-ray, 24~$\mu$m, and 1.4~GHz selected sources discussed in the previous 
subsections.  The $F_{3.6}/F_{Ks}$ colors of all three populations lie in a 
well-defined area of the color-magnitude space.  The distributions are
determined by the sensitivities at X-ray, 24~$\mu$m and 1.4~GHz, plus loose correlations 
between luminosities at these wavebands (tracing star formation and black hole accretion) 
and the rest-frame 1--4~$\mu$m luminosities (tracing stellar mass).  
At bright magnitudes the sources are at low redshifts (the blue branch), while
at the fainter fluxes the sources are at $z>0.7$ and the sources move to the red
branch. The trend is the weakest for X-ray AGNs (open squares in 
Figure~\ref{fig_color_mag-2}a), but it is still observable.  

The distributions in the $F_{4.5}/F_{3.6}$ and $F_{5.8}/F_{4.5}$ colors of 
these three populations are 
more or less similar, and we do not show them here.  
However, the $F_{8.0}/F_{5.8}$
color-magnitude diagrams become significantly different.  
A group of red $F_{8.0}/F_{5.8}$ 
sources appear at the bright end.  These colors are produced by
strong PAH emission from low-redshift starbursting 
galaxies. They are also seen in the general 8~$\mu$m selected 
diagram (Figure~\ref{fig_color_mag-1}d).
It is interesting to note the difference between the low-luminosity X-ray sources 
(X-ray luminosity $<10^{42}$~erg~s$^{-1}$, dots in Figure~\ref{fig_color_mag-2}b) 
and the X-ray AGNs 
($>10^{42}$~erg~s$^{-1}$, squares in Figure~\ref{fig_color_mag-2}b).  
In the \emph{Chandra} 2~Ms sample, none of the X-ray AGNs have red 
$F_{8.0}/F_{5.8}$ colors of $>0.8$, because the strong radiation field of 
AGNs can destroy PAH molecules.

\begin{figure}[ht!]
\epsscale{0.85}
\plotone{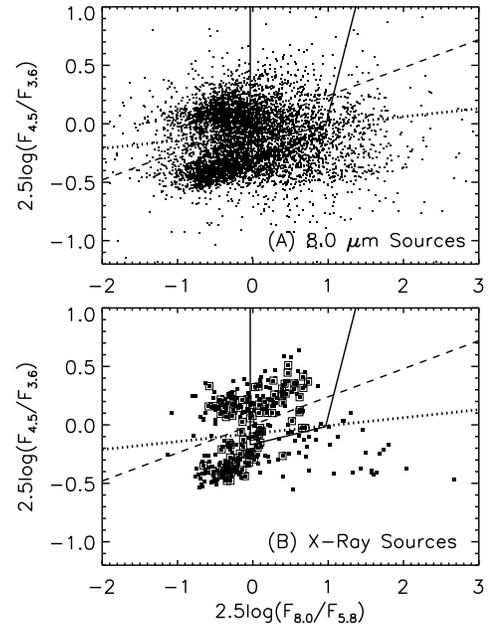}
\caption{$F_{4.5}/F_{3.6}$ vs.\ $F_{8.0}/F_{5.8}$ color-color diagrams of 
8.0~$\mu$m ($>3\sigma$)
and X-ray selected sources. In panel (B), X-ray AGNs ($>10^{42}$~erg~s$^{-1}$) 
are enclosed with
\emph{open squares}.  The \emph{dashed lines} show the selection of $z>1.3$ galaxies 
proposed by B08, which is $\log(F_{4.5}/F_{3.6})>0.096 \log(F_{8.0}/F_{5.8})$.  
The \emph{dotted lines} show the selection of $z>1.2$ galaxies proposed by 
\citet{devlin09}.  The \emph{solid lines} enclose the region where \citet{stern05} 
suggested that these colors select most broad-line AGNs.
\label{fig_color_color1}}
\end{figure}

\subsection{IRAC Color-Color Diagrams}

In B08 we presented $F_{4.5}/F_{3.6}$ vs.\ $F_{8.0}/F_{5.8}$ and
$F_{5.8}/F_{3.6}$ vs.\ $F_{8.0}/F_{4.5}$ color-color diagrams along with our spectroscopic
sample.  With these two IRAC color-color diagrams, we analyzed efficiencies
and contaminations of selections of $z>1.3$ and $z>1.6$ galaxies as well as AGNs.
Since the quality of the IRAC photometry is improved here, it is interesting to revisit 
these issues.  Here we focus on the selections of high-redshift galaxies and X-ray AGNs.  
(B08 also included the most powerful radio sources, but that sample size is very small.)

Figure~\ref{fig_color_color1} shows $F_{4.5}/F_{3.6}$ vs.\ $F_{8.0}/F_{5.8}$ color-color
diagrams of 8.0~$\mu$m and X-ray selected sources.  The color bimodality shown in
Figure~\ref{fig_color_mag-1}b is clearly seen here.
The \emph{dashed lines} are the $z>1.3$ selection proposed by B08.  
We reported in B08 that this selection can pick up 90\%  of $z>1.3$ galaxies, but
39\% of the galaxies that satisfy this criterion are $z<1.3$ contaminants.  
In the 3$\sigma$ 8.0~$\mu$m
sample here, there are 2096 spectroscopic redshifts, 352 of which are at $z>1.3$.
The above selection picks up 241 of them, corresponding to a selection
completeness of 68\%, much lower than that in B08.  On the other hand, the selection 
only picks up 54 $z<1.3$ galaxies, corresponding to a selection contamination of 18\%, also 
much lower than that in B08.  The results do not change much if we limit to higher S/N
8~$\mu$m sources where the measurements of colors are more robust.  
The inconsistency between the results here and our previous results is most likely 
a systematic difference in the calibration of the IRAC fluxes.  For example, if we 
make the galaxies here $\sim0.15$ magnitude redder in the $F_{4.5}/F_{3.6}$ color, 
then the completeness and contamination both become 
similar to those in B08.

The above comparison shows that using IRAC colors to select high-redshift galaxies
is extremely sensitive to how galactic colors are measured.  This also emphasizes that
simply adopting existing IRAC color selection schemes may be problematic unless one
is very careful about the fine details of the color measurements.
Bearing this in mind, the selection completeness and contamination are then
just a matter of fine tuning the selection criteria at the expense of one another.
Such fine tuning also requires a relatively large sample of spectroscopic redshifts.
In addition to the very likely difference in the colors, the basic conclusions about 
IRAC color selections on high-redshift galaxies remain unchanged here and in B08.  

There are also other high-redshift galaxy selections.  The \emph{dotted lines} in 
Figure~\ref{fig_color_color1} show the $z>1.2$ selection adopted by \citet{devlin09}
and \citet{marsden09} for studies of submillimeter galaxies.  We found that all of the
above conclusions apply to this selection as well.  This selection has either
incompleteness at high redshift or contaminations at low redshift.

\begin{figure}[ht!]
\epsscale{1.0}
\plotone{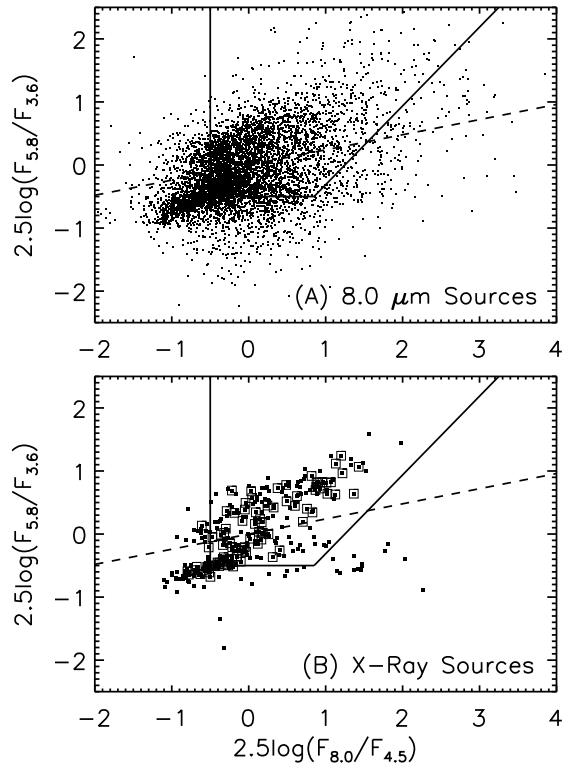}
\caption{$F_{5.8}/F_{3.6}$ vs.\ $F_{8.0}/F_{4.5}$ color-color diagrams of 
8.0~$\mu$m ($>3\sigma$)
and X-ray selected sources. In panel (B), X-ray AGNs ($>10^{42}$~erg~s$^{-1}$) 
are enclosed with
\emph{open squares}.  The \emph{dashed lines} show the selection of $z>1.6$ 
galaxies proposed by B08, which is $\log(F_{5.8}/F_{3.6})>0.096 \log(F_{8.0}/F_{4.5})$.  
The \emph{solid lines} enclose the region where \citet{lacy04} suggested
that these colors select AGNs.
\label{fig_color_color2}}
\end{figure}

Figure~\ref{fig_color_color1}b shows X-ray sources and X-ray AGNs.
The IRAC colors suggested by \citet{stern05} to select most broad-line AGNs 
are enclosed by \emph{solid lines}.  However, here we see that the Stern et al.\ 
selection only picks up a fraction of the X-ray AGNs.
By comparing both panels in Figure~\ref{fig_color_color1}
as well as Figures~\ref{fig_z_color1}b and d,  we see that the same region also selects a 
larger number of high-redshift galaxies.  These galaxies do not show up in the relatively
shallow data of \citet[their Figure~1]{stern05}.  This is consistent with the finding 
in B08 and \citet{yun08} that high-redshift reddened objects occupy similar color 
spaces to AGNs.  The selection of \citet{stern05} does not 
pick up the majority of AGNs and has substantial galaxy contamination.

B08 also presented $F_{5.8}/F_{3.6}$ vs.\ $F_{8.0}/F_{4.5}$ color-color diagrams and
used them to analyze selections of $z>1.6$ galaxies and the \citet{lacy04} selection of AGNs.
As with the $F_{4.5}/F_{3.6}$ vs.\ $F_{8.0}/F_{5.8}$ color-color diagram,
B08 concluded that the IRAC selection of $z>1.6$ galaxies can be very complete ($\sim90\%$),
but it has a large contamination ($\sim40\%$) from low-redshift galaxies.  B08 also found that
the \citet{lacy04} selection of AGNs is incomplete (though better than the
\citealp{stern05} selection) and has severe galaxy contamination.  
We present the updated $F_{5.8}/F_{3.6}$ vs.\ $F_{8.0}/F_{4.5}$ color-color diagrams in
Figure~\ref{fig_color_color2}, and we confirm all the conclusions made by B08.

To sum up, using IRAC colors to select high-redshift galaxies can be either incomplete
or suffer from contaminations from AGNs or low-redshift galaxies.  The selection can be
fine tuned to balance the completeness and contamination, but this requires a good 
spectroscopic sample and/or extremely well measured IRAC colors.  We also found that
galaxies and AGNs occupy a very similar region of color space.  
The IRAC selections of AGNs based 
on the \citet{lacy04} and \citet{stern05} methods always include large numbers of galaxy 
contaminations and have various degrees of incompleteness.

\subsection{$K_S$ and IRAC Color-Color Diagrams}

\begin{figure}
\epsscale{1.0}
\plotone{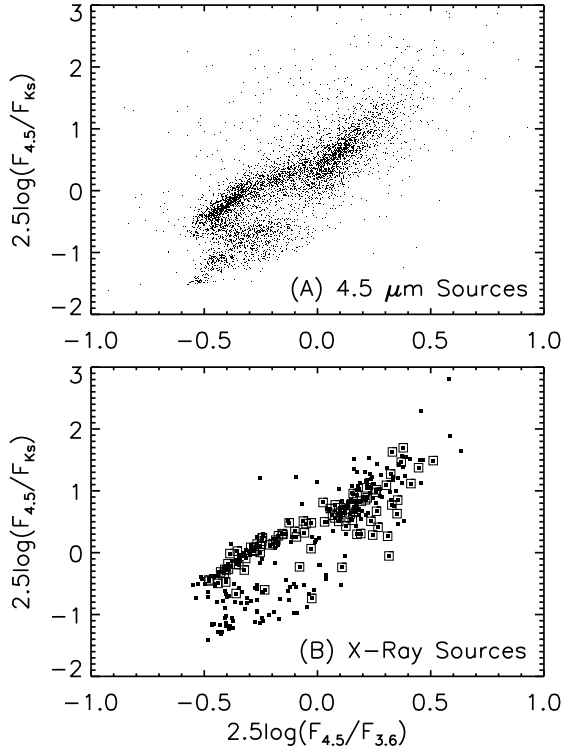}
\caption{$F_{4.5}/F_{Ks}$ vs.\ $F_{4.5}/F_{3.6}$ color-color diagrams of 
4.5~$\mu$m ($F_{4.5}>2$~$\mu$Jy)
and X-ray selected sources. In panel (B), X-ray AGNs ($>10^{42}$~erg~s$^{-1}$) 
are enclosed with \emph{open squares}. 
\label{fig_color_color3}}
\end{figure}

The above discussion only includes IRAC colors.  In this work, 
we presented ultradeep $K_S$ photometry, which can be a powerful 
tool for studying galactic colors.  We show the
$F_{4.5}/F_{Ks}$ vs.\ $F_{4.5}/F_{3.6}$ color-color diagrams for 
4.5~$\mu$m and X-ray sources in Figure~\ref{fig_color_color3}.  
Here we only adopt the two shortest IRAC bands for their higher S/N
and for their continuous availability in the \emph{Spitzer}
warm phase mission.  Again, we see that the X-ray AGNs occupy a 
color space that is nearly identical to that of galaxies.  Such a diagram 
is extremely ineffective for separating galaxies and AGNs.  On the 
other hand, this color-color diagram is much more structured than 
the previous IRAC-only diagrams, suggesting that these colors can 
provide an excellent redshift diagnostic.

\begin{figure}[h!]
\epsscale{1.0}
\plotone{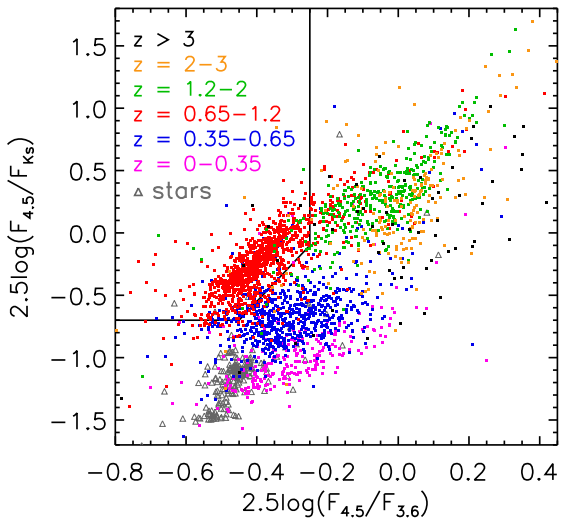}
\caption{$F_{4.5}/F_{Ks}$ vs.\ $F_{4.5}/F_{3.6}$ color-color diagram of spectroscopically
identified sources in the B08 sample.  Symbols are color coded with spectroscopic redshifts,
including spectroscopically identified stars (\emph{gray triangles}).
\label{fig_color_color3-2}}
\end{figure}

Figure~\ref{fig_color_color3-2} shows the same diagram as 
Figure~\ref{fig_color_color3}, but only for the spectroscopic 
subsample drawn from B08.  There is a very subtle difference
in the distributions of galaxies in these two diagrams 
(4.5~$\mu$m selected vs.\ spectroscopically identified), because 
the spectroscopic identifications suffer from the redshift desert and
are somewhat biased to galaxies 
with strong emission lines that can have unusual $F_{4.5}/F_{Ks}$
colors (see \S~\ref{sec_compare_sed}).  However, three distinct
redshift groups can be seen in both diagrams: 
$z<0.65$, $z=0.65$--1.2, and $z>1.2$.
In particular, we find that the following selection criteria are 
especially effective in selecting intermediate redshifts of 
$z=0.65$--1.2: 
\begin{eqnarray}
\log(F_{4.5}/F_{Ks}) > -0.28 \quad\quad {\rm and} \nonumber \\
\log(F_{4.5}/F_{3.6}) < -0.1 \quad\quad {\rm and}  \nonumber \\
\log(F_{4.5}/F_{Ks}) > 1.2 \log(F_{4.5}/F_{3.6}) + 0.26. \nonumber 
\end{eqnarray}
This selection is 
shown as the \emph{solid lines} in Figure~\ref{fig_color_color3-2}.

In Figure~\ref{fig_color_color3-2} a total of 2834 spectroscopic 
identifications are included, among which 1079 are at $z=0.65$--1.2.  
The above $K_S$ and IRAC selection picks up 870 galaxies at 
$z=0.65$--1.2, corresponding to a selection completeness of 81\%.
On the other hand, the selection also picks up 86 galaxies at 
$z<0.65$, 26 galaxies at $z>1.2$, and 3 stars, corresponding to 
a contamination of 11.6\%.  It is possible to fine tune the selection 
criteria to achieve higher completeness at the expense of increasing 
contamination, or vice versa.  For example, after we moved the 
selecting region in the diagram towards the lower-right by 0.1~magnitude 
along both the $x$ and $y$ axes, the selection completeness increased 
to 89\%, but the contamination also increased to 17\%.

$F_{4.5}/F_{Ks}$ and $F_{4.5}/F_{3.6}$ colors can still be measured 
in the \emph{Spitzer} warm phase mission 
and can be used in different ways.  The above $z=0.65$--1.2 
selection picks up galaxies at redshifts just lower than that 
in the popular $BzK$ selection 
($z\sim1.4$--2.5, \citealp{daddi04}).  Galaxies at $z>1.2$
and $z<0.65$ can also be selected from the same diagram.  
In addition, in a subsequent paper, we will use the same 
diagram to select $z>2$ massive and dusty galaxies
(Wang et al., in preparation).

Generally speaking, the addition of the $K_S$-band data increases 
the redshift resolution at $z<2$.  At higher S/N this will greatly 
help IRAC-based photometric redshift studies \citep[e.g.,][]{pope06}, 
which can be more powerful than simple color selections.
On the other hand, in all the examples discussed in this paper, 
IRAC colors become much less efficient in separating galaxies at 
$z>2$, even after the inclusion of $K_S$. For these high-redshift 
galaxies, optical colors or photometric redshifts that combine
optical, NIR, and IRAC bands \citep[e.g.,][]{mancini09} may be 
much better.  However, on optically faint objects such as high-redshift 
submillimeter galaxies \citep[e.g.,][]{wang07,younger09}, colors from 
deep NIR and IRAC imaging are valuable in studying their redshifts.

\section{Summary}\label{sec_summary}

We carried out ultradeep $K_S$ band imaging in the HDFN and its flanking 
field with WIRCam on the CFHT.  The mosaic image has a size of 
$35\arcmin \times 35\arcmin$, fully covering the GOODS-N field.  With nearly
50~hr of integration, we reached a 90\% completeness limit of 1.38~$\mu$Jy
($\rm AB=23.55$) in the 0.25~deg$^2$ region, and 0.94~$\mu$Jy
($\rm AB=23.96$) in the 0.06~deg$^2$ GOODS-N IRAC region.
The faintest detected objects have $K_S$ fluxes of $\sim0.2$~$\mu$Jy 
($\rm AB = 25.65$).  The photometry is uniform across the entire field,
and the variation is well within 2\%.  In the GOODS-N region
the astrometry is accurate to $0\farcs03$ for compact objects with 
high S/N.  The $K_S$ images and 
the associated catalogs are available through the internet.

We used the $K_S$ image and catalog as priors to measure MIR magnitudes
in the \emph{Spitzer} IRAC 3.6 to 8.0~$\mu$m bands with our REALCLEAN 
method.  We verified the REALCLEAN results using
data taken in two different epochs of observations, as well as by comparing 
our colors with the colors of stars and with the colors derived from model 
SEDs of galaxies at various redshifts.  
The REALCLEAN method appears to provide reasonably good photometry even on 
galaxies that have close neighbors within $2\arcsec$.  This also provides a 
relatively clean way to identify faint IRAC objects that are not detected in the 
$K_S$ image.  We make the REALCLEAN IRAC catalogs for both $K_S$ detected and 
undetected objects available to the community.  Additional to the REALCLEAN method,
we also measured galactic colors by convolving the $K_S$ image with the IRAC
PSFs and convolving the IRAC images with the $K_S$ PSF.  The derived colors are
consistent with the REALCLEAN results.

We found our $K_S$ sample to be extremely powerful in studying the multiwavelength 
properties of high-redshift galaxies.  The $K_S$ image detected most (if not all) of 
the X-ray, 24~$\mu$m, and radio sources published in this field.  
On the other hand, only 40\% of the \emph{HST} ACS sources in the GOODS-N 
are detected at $K_S$ at the current depth.  

We studied various $K_S$ and IRAC color-magnitude
and color-color diagrams.  Because of the rest-frame 1.6~$\mu$m bump in the SEDs of
galaxies, galaxies show significant color evolution in the $K_S$ and IRAC bands
from $z=0$ to $z<2$.  We found that using IRAC colors to select galaxies at certain
redshifts is extremely sensitive to the color measurements and that balancing 
selection completeness and contamination will require careful tuning of the selection
criteria with a good spectroscopic sample.  We discussed an effective selection
of $z=0.65$--1.2 galaxies using the $K_S$ band and the two shorter IRAC bands that
will remain available in the \emph{Spitzer} warm phase mission.  We also confirmed our 
previous studies (B08) that the IRAC AGN selections in \citet{lacy04} and \citet{stern05} 
are either incomplete at the low-luminosity end or suffer severely from galaxy contamination.

\acknowledgments
We thank the referee for comments that greatly improve the manuscript.
We are grateful to the CFHT staff for help on obtaining the data, C.-H.\ Yan for 
providing useful parameters about WIRCam, and B.-C.\ Hsieh and L.\ Lin for 
useful discussions.  
This work started when W.H.W. was a Jansky Fellow at the 
National Radio Astronomy Observatory.  
We gratefully acknowledge support
from NRAO (W.H.W.), the National Science Council of Taiwan 
grant 98-2112-M-001-003-MY2 (W.H.W.), 
NSF grants AST 0709356 (L.L.C.) and AST 0708793 (A.J.B.), the
University of Wisconsin Research Committee with funds granted by the 
Wisconsin Alumni Research Foundation (A.J.B.), and the David and 
Lucile Packard Foundation (A.J.B.).




















\end{document}